\documentclass[11pt,oneside]{article}

\usepackage{amsthm}
\usepackage{amsmath,amssymb,amsfonts}

\usepackage{graphicx}

\usepackage{color}
\usepackage[dvipsnames]{xcolor}
\usepackage{natbib}
\usepackage{fullpage, url, graphicx, subfigure}
\usepackage{fancyhdr}
\usepackage{mathrsfs}
\usepackage{epsfig}
\usepackage{graphics}
\usepackage{epsf}
\usepackage{amsmath, amsthm, amssymb, multirow, paralist, mathtools}
\usepackage{algorithm,algorithmic}
\usepackage{tikz}
\usepackage{color}
\usepackage{caption}
\usepackage{lscape}
\usepackage{array}
\usepackage{epstopdf}

\def\dif{\mathrm d}

\def\T{ {\mathrm{\scriptscriptstyle T}} }

\def\TV{\mbox{TV}}

\newtheorem{prop}{Proposition}

\newtheorem{definition}{Definition}
\newtheorem{eg}{Example}

\begin{document}

\begin{titlepage}

\begin{center}
{\Large Hierarchical Total Variations and Doubly Penalized ANOVA Modeling for Multivariate Nonparametric Regression}

\vspace{.15in} Ting Yang\footnotemark[1] and Zhiqiang Tan\footnotemark[1]

\vspace{.1in}
\today
\end{center}

\footnotetext[1]{Ting Yang is with Yelp, Inc., San Francisco, CA 94105, and Zhiqiang Tan is Professor, Department of Statistics, Rutgers University,
Piscataway, NJ 08854 (E-mail: ztan@stat.rutgers.edu). }

\paragraph{Abstract.}

For multivariate nonparametric regression, functional analysis-of-variance (ANOVA) modeling aims to capture the relationship between a response and covariates
by decomposing the unknown function into various components,  representing main effects, two-way interactions, etc.
Such an approach has been pursued explicitly in smoothing spline ANOVA modeling and implicitly in various greedy methods such as MARS.\
We develop a new method for functional ANOVA modeling, based on doubly penalized estimation using total-variation and empirical-norm penalties,
to achieve sparse selection of component functions and their knots.
For this purpose, we formulate a new class of hierarchical total variations, which measures total variations at different levels including main effects and multi-way interactions,
possibly after some order of differentiation.
Furthermore, we derive suitable basis functions for multivariate splines such that the hierarchical total variation can be represented as a regular Lasso penalty,
and hence we extend a previous backfitting algorithm to handle doubly penalized estimation for ANOVA modeling.
We present extensive numerical experiments on simulations and real data to compare our method with existing methods including MARS, tree boosting, and random forest.
The results are very encouraging and demonstrate considerable gains from our method in both prediction or classification accuracy and simplicity of the fitted functions.

%\vspace{-.2in}

\end{titlepage}

\section{Introduction} \label{sec:intro}

A fundamental problem in statistics and related fields is multivariate nonparametric regression, that is,
to estimate a nonparametric mean function from a collection of independent observations of a response variable and covariates, denoted as
$Y_i$ and $X_i=(X_{i1}, \ldots, X_{ip})$ for $i=1,\ldots, n$.
For continuous responses, nonparametric regression can be defined such that
\begin{equation}\label{eqn:model}
  Y_i = f(X_{i1},  \ldots, X_{ip}) + \varepsilon_i,
\end{equation}
where $f(x)=f(x_1,\ldots,x_p)$ is an unknown function,
and $\varepsilon_i$ is a noise with mean zero and a finite variance given $X_i$.
In general, the objective is to construct an estimator $\hat f(x )$ achieving accurate approximation to $f(x )$ over
a flexible class of functions.

Consider the framework of functional analysis of variance (ANOVA) modeling, in which the multivariate function $f$ is decomposed as
\begin{align}\label{eqn:f-anova}
  f(x_1, \ldots, x_p) &= f_0 + \sum_{1 \leq j_1 \leq p}f_{j_1}(x_{j_1}) + \sum_{1 \leq j_1 < j_2 \leq p}f_{j_1, j_2}(x_{j_1}, x_{j_2}) + \cdots +  \nonumber  \\
  & \quad \sum_{1\leq j_1 <\cdots < j_K \leq p}f_{j_1,\ldots, j_K}(x_{j_1}, \cdots, x_{j_K}),
\end{align}
where $f_0$ is a constant, $f_{j_1}$'s are univariate functions representing main effects,
$f_{j_1, j_2}$'s are bivariate functions representing two-way interactions, etc, and $K$ is the maximum way of interactions allowed.
For identifiability, the individual functions in (\ref{eqn:f-anova}) are required to
satisfy side conditions similarly as in parametric ANOVA.
Notably, this framework has been used to develop smoothing spline ANOVA modeling \citep{wahba1995smoothing,gu2002penalized}, where the functions in (\ref{eqn:f-anova}) are assumed
to lie in tensor-product reproducing kernel Hilbert spaces (RKHSs), particularly those defined from univariate Sobolev-$L_2$ spaces associated with smoothing splines.
For estimation, it is common to use penalized least squares, where the penalty is a sum of squared RKHS norms of the component functions in (\ref{eqn:f-anova}). Alternatively,
the penalty can be a sum of RKHS norms to mimic Lasso \citep{tibshirani1996regression}, as proposed by \cite{lin2006component}.

The representation (\ref{eqn:f-anova}) with $K=1$ leads to standard additive models,
$f(x )=f_0 + \sum_{j=1}^p f_j(x_j)$, where $f_j$'s are flexible univariate functions \citep{stone1986dimensionality}.
Theory and methods for additive modeling has been extensively studied in classical settings with $p$ much less than $n$ \citep{hastie1990generalized,wood2017generalized} and
recently high-dimensional settings with $p$ close to or greater than $n$ \cite[e.g.,][]{ravikumar2007spam,meier2009high,koltchinskii2010sparsity,raskutti2012minimax,petersen2016fused,yang2018backfitting,tan2018doubly}.
One of the important ideas from high-dimensional additive modeling is to use {\it doubly penalized estimation},
where the estimators $\hat f_0$ and $(\hat f_1,\ldots,\hat f_p)$ are defined as a minimizer of
\begin{equation} \label{eqn:doubly-pen}
\frac{1}{2} \|Y - f_0 - \sum_{j=1}^p f_j \|_n^2  + \sum_{j=1}^p \{ \rho \| f_j \|_F + \lambda \|f_j \|_n \},
\end{equation}
over a constant $f_0$ and univariate functions $(f_1,\ldots, f_p)$ for some tuning parameters $(\rho,\lambda)$.
Here $\|\cdot\|_n$ is the empirical $L_2$ norm based on the data points, e.g.,
$\|f_j \|_n = \{n^{-1} \sum_{i=1}^n f_j^2(X_{ij}) \}^{1/2}$,
and $\| f_j \|_F$ is a functional semi-norm describing the complexity of $f_j$.
For $m \ge 1$, denote by $D^m f_j$ the $m$th derivative of $f_j$.
Examples of semi-norms include the Sobolev-$L_2$ semi-norm of order $m$, defined as $\| D^m f_j \|_{L_2} = \{ \int (D^m f_j(z))^2 \,\dif z \}^{1/2}$,
and the $m$th-order total variation, defined as $\TV(D^{m-1}f_j)$,
where for a univariate function $g_j$ on an interval $[a_j,b_j]$,
$$
\TV(g_j) = \sup \left\{\sum_{i=1}^k |g_j (z_{i})- g_j(z_{i-1})|: a_j \le z_0 < z_1<\ldots<z_k \le b_j \mbox{ for any } k \ge 1 \right\}.
$$
If $g_j$ is differentiable with derivative $D g_j$, then $\TV(g_j) = \int | D g_j(z)| \,\dif z$.

The two penalties on each function $f_j$ in (\ref{eqn:doubly-pen}) serve distinct but complementary roles in determining a solution for $f_j$,
as discussed in \cite{yang2018backfitting}.
First, the functional semi-norm $\| f_j\|_F$ is used to induce smoothness for $f_j$.
In particular, for the $m$th total-variation penalty, if $f_j$ is restricted to be a spline of order $m$
with possible knots in $\{X_{ij}: i=1,\ldots,n\}$, i.e., a piecewise polynomial of degree $m-1$ and $(m-2)$th continuously differentiable,
then minimization of (\ref{eqn:doubly-pen}) can yield a solution $\hat f_j$ with only a few knots from the data points, hence achieving a sparse selection of knots
similarly as in \cite{osborne1998knot} for univariate smoothing.
Second, the empirical norm $\|f_j\|_n$ is used to induce sparsity for $f_j$. An entirely zero solution $\hat f_j$ can be obtained via soft thresholding similarly as how
zero coefficients can be obtained for some regressors in linear regression with the Lasso penalty.

In this article, we develop a new method for functional ANOVA modeling based on doubly penalized estimation, to achieve sparse selection of component functions and their knots.
Conceptually, the ANOVA  representation (\ref{eqn:f-anova}) can be seen as an extended additive model with potentially a large number of component functions,
and this motivates the use of sparsity-driven techniques.
In general, a doubly penalized loss function is of the form
\begin{equation} \label{eqn:doubly-pen2}
\frac{1}{2} \|Y - f \|_n^2  + \sum_{k=1}^K \sum_{1\leq j_1 <\cdots < j_k \leq p} \{ \rho_k \| f_{j_1,\ldots, j_k}  \|_F + \lambda_k \|f_{j_1,\ldots, j_k}\|_n \} ,
\end{equation}
where $\|\cdot\|$ is the empirical $L_2$ norm as above, $\|\cdot\|_F$ is a functional semi-norm to be specified, and $(\rho_k,\lambda_k)$ are tuning parameters.
Our work involves two main contributions.
\begin{itemize}\addtolength{\itemsep}{-.05in}
\item We construct a new class of total-variation penalties, called hierarchical total variations, which
measures total variations at different levels including main effects and multi-way interactions for multivariate functions,
for a fixed order of differentiation $m$.

\item We derive suitable basis functions for multivariate splines, defined as a tensor product of univariate spline spaces,
such that each penalty $\| f_{j_1,\ldots, j_k}  \|_F$ is reduced to a
regular Lasso penalty. Then we extend the backfitting algorithm of \cite{yang2018backfitting}, called block descent and thresholding,
to numerically minimize (\ref{eqn:doubly-pen2}).
\end{itemize}
Similarly as mentioned earlier, our use of double penalties implies that (i) each fitted component $\hat f_{j_1,\ldots, j_k}$ can be entirely zero,
and (ii) if nonzero, $\hat f_{j_1,\ldots, j_k}$ is
piecewise constant or piecewise cross-linear for differentiation order $m=1$ or 2, potentially with a few knots selected from a large set of $k$-dimensional knots.
Throughout, a multivariate function is said to be piecewise cross-linear if it is piecewise linear in each coordinate with all other coordinates being fixed.

The rest of the paper is organized as follows. Section~\ref{sec:related} reviews related work. In Section~\ref{sec:method},
we discuss the new method for functional ANOVA modeling with linear and logistic links for continuous and binary responses.
Section~\ref{sec:experiment} present numerical experiments on simulated and real datasets. Section~\ref{sec:conclusion} concludes the paper.
All proofs are provided in the Appendix.

\section{Related work}\label{sec:related}

There is a vast literature on multivariate nonparametric regression. For space limitation, we only discuss directly related work to ours.

{\bf ANOVA modeling.} As mentioned earlier, standard additive modeling with $K=1$ in (\ref{eqn:f-anova}) has been extensively studied in low- and high-dimensional settings
for the sake of dimension reduction by ignoring all possible interactions.
Although a broader view of additive modeling could accommodate ANOVA modeling by allowing multivariate component functions, practical methods for
functional ANOVA modeling with interactions have been mainly developed in the smoothing spline approach, which heavily draws on theory of tensor-product RKHSs \citep{wahba1995smoothing,gu2002penalized,lin2006component}.
For estimation, penalized least squares and maximum likelihood in these existing methods only involve roughness penalties defined from RKHS norms.
By comparison, we pursue a distinctly different approach, which not only involves total-variation penalties carefully extended to multivariate functions, but also
employs doubly penalized estimation incorporating empirical-norm penalties.
Our approach can be readily modified to develop doubly penalized ANOVA modeling using RKHS-norm and empirical-norm penalties, which would lead to sparse selection of
component functions but not that of their knots.
Moreover, use of total-variation penalties is known to achieve faster convergence rates in the presence of spatially inhomogeneous smoothness than that of smoothing-spline type penalties,
at least in univariate smoothing \citep{donoho1994ideal,mammen1997locally}.
We expect that similar results can also be obtained in the context of ANOVA modeling, but leave such theoretical investigation to future work.

Other related methods in the spirit of ANOVA modeling includes \cite{choi-li-zhu2010}, who studied regression models with fixed regressors and two-way interactions, and
\cite{Radchenko-James2010}, who considered ANOVA model (\ref{eqn:f-anova}) with nonlinear $f_{j_1}$'s and $f_{j_1,j_2}$'s for $K=2$
but, compared with our method, employed a penalty depending only on empirical norms of the component functions.
These two methods are motivated to satisfy the condition that an interaction term can be added only if the two corresponding main effects are also included.
This heredity condition can be appealing in terms of variable or component selection, but its effect on prediction accuracy may depend on applications.
It will be of interest to extend our method using similar ideas for achieving the heredity condition.

{\bf Greedy methods.} The fact that each fitted component $\hat f_{j_1,\ldots, j_k}$ and hence the overall function $\hat f$ in our method
is piecewise constant or cross-linear, depending on $m=1$ or 2, indicates interesting connections to a number of
greedy methods based trees or splines, such as %CART (Brieman et al. 1984),
MARS \citep{friedman1991multivariate}, tree boosting \citep{freund1997decision,trevor2009elements}, and random forests \citep{breiman2001random}.,
For these methods, the underlying model is of the form
\begin{align}
f(x) = \gamma_0 + \sum_{j=1}^J \gamma_j h_j(x),
\end{align}
where $h_j(x)$ is a base function and $\gamma_j$ its associated coefficient. For MARS, $h_j(x)$ is a truncated-linear basis function or a product of such functions and hence
$f(x)$ is piecewise cross-linear. For tree boosting or random forests, $h_j(x)$ is a tree with constant values over hyper-rectangles and hence $f(x)$ is piecewise constant.
These greedy methods are designed such that the parameters associated with each base function $h_j(x)$ including split values, the node size, and the coefficient $\gamma_j$
are determined in an iterative manner by forward selection and sometimes backward deletion or modification. Regularization is handled through the number of iterations $J$, known as early stopping, and
other training techniques such as shrinkage and subsampling, instead of explicit penalty terms as in our method.
Although there is a close relationship between boosting with small step-sizes and Lasso as shown in linear regression by \cite{trevor2009elements} Section 16.2,
comparison of the two approaches, greedy approximation and penalized estimation, in complex settings remains to be fully understood.
Compared with MARS, tree boosting, and random forest in numerical experiments, our method is found to not only achieve competitive or superior prediction or classification performance
but also yield simpler fitted functions with greater sparsity and hence easier interpretation.

{\bf Total variations.} Use of total-variation penalties has often been restricted to univariate and bivariate functions in statistics and computer science, although
mathematical theory of multivariate functions of bounded variations seems to be well studied \citep{ambrosio2000functions}.
For univariate smoothing, \cite{mammen1997locally} studied regression splines using total-variation penalties with an arbitrary differentiation order $m$.
A closely related method is trend filtering \citep{kim2009ell_1,tibshirani2014adaptive}.
Additive modeling has also been pursued using total variations on univariate functions in low- and high-dimensional settings
\cite[e.g.,][]{petersen2016convex,sadhanala2017additive,yang2018backfitting}.

There are various types of total variations for bivariate functions.
Suppose that the design points are on a 2D lattice, that is, $\{ (X_{i1}, X_{i2}): i=1,\ldots, n\} = \{(z_{i1}, z_{j2}): i=1,\ldots, n_1, j=1,\ldots,n_2\}$.
\cite{mammen1997locally} studied penalized least squares with a total-variation penalty on a bivariate function $g(x_1,x_2)$ as follows,
\begin{equation*}
\mathrm{ATV}_2^1 (g; \rho_1, \rho_2) = \rho_2 \mathrm{TV}_2 (g) + \rho_1\Big(\mathrm{TV}_1 (g_1) + \mathrm{TV}_1 (g_2) \Big),
\end{equation*}
where $g_1(x_1)= \sum_{j=1}^{n_2} g(x_1, z_{j2})/n_2$, $g_2(x_2) = \sum_{i=1}^{n_1}f(z_{i1}, x_2)/n_1$, and $\mathrm{TV}_1$ and $\mathrm{TV}_2$ are
as in Definition~\ref{def:multivariate-tv} with $d=1$ and $2$.
In fact, $\mathrm{ATV}_2^1$ is a special case of Definition \ref{def:generalized-total-variation} with $d=2$ and $m=1$, suitable for piecewise constant functions.
We extend this idea and carefully construct a class of total-variation penalties for multivariate functions of three or more variables while allowing
an arbitrary order of differentiation $m$.
Moreover, as another extension, we employ total-variation penalties on component functions in ANOVA modeling, where the design points, in general, do not constitute a lattice in any dimensions.

Another type of total variations is widely used in image denoising, where a bivariate function $g(x_1,x_2)$ represents an image of size $n_1 \times n_2$ \citep{rudin1992nonlinear}.
The isotropic total variation is defined by discretizing the 2D total variation from mathematical analysis:
\begin{equation*}
\sum_{i, j} \Big( | g(z_{i+1,1}, z_{j2}) - g(z_{i1}, z_{j2})|^2 + |g(z_{i1}, z_{j+1,2}) - g(z_{i1}, z_{j2}) |^2 \Big)^{1/2},
\end{equation*}
where a finite difference across a boundary, such as $g(z_{n_1+1,1},z_{j2})- g(z_{n_1,1},z_{j2})$ is assumed to be zero. A simple anisotropic total variation is defined as
\begin{equation*}
\sum_{i, j} \Big( | g(z_{i+1,1}, z_{j2}) - g(z_{i1}, z_{j2})| + |g(z_{i1}, z_{j+1,2}) - g(z_{i1}, z_{j2}) | \Big) .
\end{equation*}
These total variations and other versions \cite[e.g.,][]{condat2017discrete} are different from our class of total variations for ANOVA modeling, including $\mathrm{TV}_2$ or $\mathrm{ATV}_2^1$ above.

For bivariate regression,  \cite{petersen2016fused} proposed a penalized method, which fits a piecewise constant model by penalized least squares with a group Lasso penalty.
This penalty is total-variation like, measuring the differences between neighboring rows and columns of the image represented by the bivariate function, but differs from our construction of total variations. An extension of the method was then presented for fitting an additive model of bivariate functions, although no numerical experiments were reported.

%Freund, Y. and R. Schapire (1997). A decision-theoretic generalization of on-line learning and an application to boosting. Journal of Computer and System Sciences 55(1), 119–139.

%L. Ambrosio, N. Fusco, D. Pallara, "Functions of bounded variations and free discontinuity problems". Oxford Mathematical Monographs. Oxford University Press, New York, 2000

%L. Rudin, S. Osher, and E. Fatemi, Nonlinear total variation based noise removal algorithms, Physica D, 60 (1992), pp. 259-268.

%Ashley Petersen, Noah Simon, Daniela Witten (2016), Convex Regression with Interpretable Sharp Partitions, Journal of Machine Learning Research, 17 (2016) 1-31.

\section{Method}\label{sec:method}

We develop a doubly penalized method for ANOVA modeling, using empirical-norm and total-variation penalties, to achieve both smoothness and sparsity in component functions.
One of the main challenges is to construct an appropriate class of total-variation penalties, which not only measures total variations hierarchically from main effects to multi-way interactions,
but also facilitates representation of such penalties in the form of regular Lasso penalties on coefficients associated with suitable basis functions.

\subsection{Hierarchical total variations} \label{sec:HTV}

First, we define a raw total variation, measuring $d$-way interactions for $d$-variate functions.
Let $g(z) = g(z_1, \ldots, z_d)$ be a $d$-variate function on a product domain $\overline{\mathcal Z} = \prod_{j=1}^d \overline{\mathcal Z}_{j}$,
where $z_j\in \overline{\mathcal Z}_{j}$ is the $j$th coordinate of $z\in \overline{\mathcal Z}$. Consider a $d$-dimensional grid $\mathcal Z$ in $\overline{\mathcal Z}$,
formed with the marginal knots $z_{1, j}<z_{2, j}<\ldots < z_{n_j, j}$ in the $j$th coordinate.

\begin{definition}\label{def:multivariate-tv}
Given grid $\mathcal Z$, the raw total variation (TV) of $g(z_1, z_2, \ldots, z_d)$ is defined as
\begin{align*}
  & \mathrm{TV}_{d}\big(g(z_1, z_2, \ldots, z_d)\big)=\sum_{i_1=1}^{n_1-1}\cdots\sum_{i_d=1}^{n_d-1}  \\
  & \Big|g(z_{i_1+1, 1}, \ldots, z_{i_d+1, d}) +
  \sum_{l=1}^d (-1)^l \sum_{1 \le j_1 < \cdots < j_l \le d} g(z_{i_1+1, 1}, \ldots, z_{i_{j_1}, j_1}, \ldots, z_{i_{j_l}, j_l}, \ldots, z_{i_d+1, d}) \Big|
\end{align*}
\end{definition}

\begin{figure}[H]
  \centering
  \includegraphics[width=5in, height=1.5in]{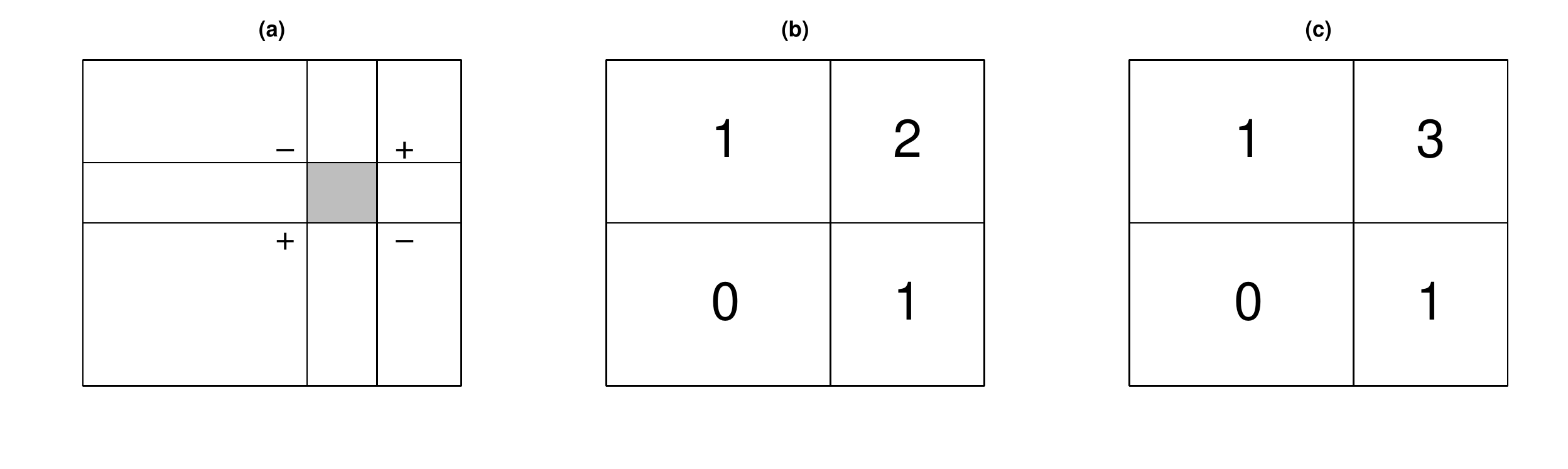}\\
  \caption{Illustration of raw total variations for bivariate functions}\label{fig:dtv-visual}
\end{figure}

For example, the total variations with $d$ from 1 to 3 are as follows:
\begin{enumerate}
  \item For $d=1$, $\mathrm{TV}_1\big(g(z_1)\big)=\sum_{i_1=1}^{n_1-1}\Big|g(z_{i_1+1, 1})-g(z_{i_1, 1})\Big|$
  \item For $d=2$, $\mathrm{TV}_2\big(g(z_1, z_2)\big)=\sum_{i_1=1}^{n_1-1}\sum_{i_2=1}^{n_2-1}\Big|g(z_{i_1+1, 1}, z_{i_2+1, 2})-g(z_{i_1, 1}, z_{i_2+1, 2})$\\
      $-g(z_{i_1+1, 1}, z_{i_2, 2})+g(z_{i_1, 1}, z_{i_2, 2})\Big|$
  \item For $d=3$, $\mathrm{TV}_3\big(g(z_1, z_2, z_3)\big)=\sum_{i_1=1}^{n_1-1}\sum_{i_2=1}^{n_2-1}\sum_{i_3=1}^{n_3-1}\Big|g(z_{i_1+1, 1}, z_{i_2+1, 2}, z_{i_3+1, 3})$
      \\
      $-g(z_{i_1+1, 1}, z_{i_2+1, 2}, z_{i_3, 3})-g(z_{i_1+1, 1}, z_{i_2, 2}, z_{i_3+1, 3})-g(z_{i_1, 1}, z_{i_2+1, 2}, z_{i_3+1, 3})$ \\
      $+g(z_{i_1, 1}, z_{i_2, 2}, z_{i_3+1, 3})+g(z_{i_1, 1}, z_{i_2+1, 2}, z_{i_3, 3})+g(z_{i_1+1, 1}, z_{i_2, 2}, z_{i_3, 3}) -g(z_{i_1, 1}, z_{i_2, 2}, z_{i_3, 3})\Big|$
\end{enumerate}
In general, the raw TV is a sum of absolute values over all individual cells in the grid $\mathcal Z$. Each absolute value measures
the magnitude of the $d$-way interaction of $g(z)$ evaluated at the corners of the corresponding cell.
The case of $d=2$ is illustrated by Figure~\ref{fig:dtv-visual}.
Two simple bivariate functions are also shown, where the raw TV is zero for the additive function (middle)
but is one for the function with a two-way interaction (right).

The raw TV of a $d$-variate function $g(z)$ measures only $d$-way interactions while ignoring any lower-way interactions including main effects. In fact,
$\TV_d(g)$ is invariant if $g(z)$ is modified by adding any function of less than $d$ variables.
To properly account for interactions at various levels, we introduce a hierarchical total variation by combining raw total variations of
projections of a $d$-variate function into margins of different dimensions.
Moreover, we formulate an extension which measures total variations after a certain order of differentiation and hence is
suitable for piecewise cross-linear functions and beyond.
%Intuitively, the projection of $g(z)$ into a margin $(z_{j_1}, \ldots, z_{j_k})$ can be defined by evaluating $g(z)$ with the complementary margin fixed at some point
%or averaging $g(z)$ over the complementary margin.
To give formal definitions, we exploit the following notations for functional ANOVA.

Let $H_j$ be a projection operator such that $H_j g$ is constant in the $j$th coordinate and $H_j^2 = H_j$. Two main examples of such operators are (i) a fixed-point operator, $F_j$, defined
such that $F_j g(z) = g(z_1, \ldots, z_j^*, \ldots, z_d)$ where $z_j^*$ is some fixed point in $\{z_{1j}, \ldots, z_{n_j,j}\}$, or (ii) an averaging operator, $A_j$, defined such that
$A_j g(z) = \sum_{i=1}^{n_j} g(z_1, \ldots, z_{ij}, \ldots, z_d)/n_j$.
Then the multi-way ANOVA decomposition of $g$ can be written as \citep{gu2002penalized}
\begin{equation}\label{eqn:anova-decomp}
  g = \Big\{\prod_{j=1}^{d}(I - H_j + H_j)\Big\}g = \sum_{k=0}^{d}\sum_{S_k:|S_k|=k} g_{S_k},
\end{equation}
where the summation is over all $2^d$ subsets $S_k\subset \{1, \ldots, d\}$ and
\begin{equation}\label{eqn:anova-fk}
  g_{S_k} = \prod_{j \in S_k}(1-H_j)\prod_{j\not\in S_k}H_jg .
\end{equation}
The term $g_{\emptyset}$ is a constant, $g_{\{j_1\}} = (I-H_{j_1})\prod_{j\neq j_1}H_j g$ is a main effect, $g_{\{j_1,j_2\}} = (I-H_{j_1})(I-H_{j_2})\prod_{j\neq j_1, j_2}H_r g$ is a two-way interaction, and so on.
In general, the term $g_{S_k}$ varies only in $\{z_j: j \in S_k\}$ and satisfies the side condition that $H_j g_{S_k} = 0$ for all $j \in S_k$.

\begin{definition}\label{def:generalized-total-variation}
Let $m \ge 1$ be a fixed order of differentiation, and $(\rho_1,\ldots,\rho_d)$ be positive constants. The hierarchical total variation is defined inductively as follows.
\begin{itemize}
  \item For $m=1$,
  \begin{eqnarray*}
\mathrm{HTV}_d^{1}\big(g; \rho_1, \ldots, \rho_d\big) = \sum_{k=1}^{d}\sum_{S_k: |S_k|=k} \rho_k \mathrm{TV}_k\Big(\prod_{\substack{j \not\in S_k}}H_{j}g \Big).
\end{eqnarray*}
  \item For $m \ge 2$,
\begin{eqnarray*}
\mathrm{HTV}_d^{m}\big(g; \rho_1, \ldots, \rho_d\big) = \sum_{k=1}^{d}\sum_{S_k: |S_k|=k} \mathrm{HTV}_k^{m-1}\Big(\prod_{j \not\in S_k}H_{j} \prod_{j \in S_k}D_{j}g; \rho_1, \ldots, \rho_k\Big),
\end{eqnarray*}
where $D_j$ is the differentiation operator with respect to $z_j$.
\end{itemize}
In particular, the hierarchical total variation is denoted as $\mathrm{FTV}_d^{m}$ or $\mathrm{ATV}_d^{m}$ if
the operator $H_j$ is specified as the fixed-point operator $F_j$ or the averaging operator $A_j$.
\end{definition}

For a univariate function $g(z_1)$, the hierarchical TV reduces to the standard definition $\mathrm{HTV}_1^{m}\big(g; \rho_1) = \rho_1 \TV_1( D_1^{m-1} g)$ \cite[e.g.,][]{mammen1997locally}.
For a bivariate function $g(z_1, z_2)$, the hierarchical TV with $m=1$ or $m=2$ is defined as
\begin{align}
\mathrm{HTV}_2^{1}\big(g; \rho_1, \rho_2\big) &=\rho_2 \mathrm{TV}_2\big(g\big) + \rho_1\Big(\mathrm{TV}_1\big(H_1g\big) + \mathrm{TV}_1\big(H_2g\big)\Big), \label{eqn:2d-tv-constant} \\
\mathrm{HTV}_2^{2}\big(g; \rho_1, \rho_2\big) &= \rho_2 \mathrm{TV}_2\big(D_1D_2g\big) + \rho_1\Big(\mathrm{TV}_1\big(H_1D_1D_2g\big) + \mathrm{TV}_1\big(H_2D_1D_2g\big)\Big) \nonumber \\
   &\quad + \rho_1\Big(\mathrm{TV}_1\big(H_1D_2g\big) + \mathrm{TV}_1\big(H_2D_1g\big)\Big) .  \label{eqn:2d-tv-linear}
\end{align}
For the operator $H_j$ specified as the averaging operator $A_j$, the hierarchical TV in (\ref{eqn:2d-tv-constant}) reduces to that used in \cite{mammen1997locally}.
This penalty is suitable for piecewise constant functions as approximations in nonparametric regression. By comparison, the hierarchical TV in (\ref{eqn:2d-tv-linear})
appears to be new and is suitable for piecewise cross-linear functions as approximations. This penalty involves not only hierarchical TV (of differentiation order 1) applied to the cross-derivative $D_1D_2 g$,
but also the raw TV applied to the ``marginalized" derivatives $H_1 D_2g$ and $H_2 D_1 g$.
See Examples~\ref{eg:constant} and \ref{eg:linear} later for further discussion.

Our construction of the hierarchical TV is designed to achieve two purposes. The first is to measure the overall complexity of interactions at various levels including main effects,
whereas the second is to facilitate a Lasso representation of such penalties in terms of coefficients of basis functions.
The second property is made explicit in Proposition~\ref{prop:lasso} later.
The first property can be illustrated by the following result, which shows how the hierarchical TV is decomposed according to various levels of interactions in the ANOVA decomposition (\ref{eqn:anova-decomp}).

\begin{prop}\label{prop:additive}
For a $d$-variate function $g(z)$ with ANOVA decomposition (\ref{eqn:anova-decomp}), the hierarchical total variation can be equivalently written as follows. For $m=1$,
\[
\mathrm{HTV}_d^{1}\big(g; \rho_1, \ldots, \rho_d\big) = \sum_{k=1}^{d}\sum_{S_k: |S_k|=k}\rho_k\mathrm{TV}_k\big(g_{S_k}\big),
\]
and for $m \ge 2$,
\[
\mathrm{HTV}_d^{m}\big(g; \rho_1, \ldots, \rho_d\big) = \sum_{k=1}^{d}\sum_{S_k: |S_k|=k}\mathrm{HTV}_k^{m-1} \Big(\prod_{j\in S_k}D_j g_{S_k}; \rho_1, \ldots, \rho_d\Big).
\]
\end{prop}

The above expressions for the hierarchical TV are informative in terms of the component functions in the ANOVA decomposition.
On the other hand, we caution that a direct use of Proposition~\ref{prop:additive} for calculating the hierarchical TV is not as simple as it may appear, especially for $m\ge 2$.
This is because calculation of $\mathrm{HTV}_k^{m-1}\big(\prod_{j\in S_k}D_jg_{S_k}\big)$ inductively by Proposition~\ref{prop:additive}
would require finding the ANOVA decomposition of $\prod_{j\in S_k}D_jg_{S_k}$, which is not immediate.

\subsection{Basis functions and Lasso representation} \label{sec:basis-transform}

In the univariate case, the standard TV for a spline with fixed knots can be represented as a Lasso penalty on the coefficients associated with truncated power basis functions \cite[e.g.,][]{yang2018backfitting}.
The hierarchical TV is much more complicated when considering multivariate functions with $d\ge 2$ instead of univariate functions.
Nevertheless, we show in this section how to
derive a Lasso representation of the hierarchical TV for a multivariate spline by carefully transforming products of truncated power basis functions.

First, we introduce the truncated power basis in each coordinate. As in Section~\ref{sec:HTV}, consider a $d$-dimensional grid $\mathcal Z$, formed with the marginal
knots $z_{1,j} < z_{2,j} < \cdots < z_{n_j,j}$ in the $j$th coordinate. For an order of differentiation $m \ge 1$, define a knot ``superset"

\begin{equation*}
\{t_{1,j},\ldots, t_{n_j-m,j}\}= \left\{
\begin{aligned}
\{z_{(m-1)/2+2,j}, \ldots, z_{n_j-(m-1)/2,j}\} & \quad \mbox{if} \ m\ \mbox{is} \ \mbox{odd}, \\
\{z_{m/2+1,j}, \ldots, z_{n_j-m/2,j}\} & \quad \mbox{if} \ m \ \mbox{is} \ \mbox{even}.\\
\end{aligned}
\right.
\end{equation*}
where the points near the left and right boundaries are removed to avoid over-parameterization \citep{mammen1997locally}.
The truncated power basis set, denoted as $\overline{\mathcal B}_j^m$, in the $j$th coordinate consists of
\begin{align*}
  & \phi^{(m)}_{\nu, j}(z_j) = z_j^{\nu-1} /(\nu-1)!, \quad \nu = 1, \ldots, m, \\
  & \phi^{(m)}_{\nu+m, j}(z_j) = (z_j -t_{\nu,j})_+^{m-1} /(m-1)!, \quad \nu = 1, 2, \ldots, n_j-m,
\end{align*}
where $(c)_+ = \max(0,c)$ and $(c)^0_+ = 0$ if $c<0$ or 1 if $c \ge 0$.
Denote $\mathcal I=\{1\}$ and $\mathcal B_j^m = \{\phi^{(m)}_{2,j}, \ldots, \phi^{(m)}_{n_j,j}\}$ such that $\overline{\mathcal B}_j^m = \mathcal I \cup \mathcal B_j^m$.
A univariate spline of order $m$ in $z_j$ can be written as $\alpha_0 + \alpha^\T_j \phi^{(m)}_j$, where
$\alpha_0$ is an intercept, $\alpha_j$ is a column vector of coefficients, and $\phi^{(m)}_j$ is a column vector of the basis functions in $\mathcal B_j^m$,
both of dimension $n_j-1$.

Next, we define a set of multivariate splines of cross-order $m$ as the tensor product of the $d$ sets of univariate splines of order $m$.
Recall that, for simplicity, a tensor product of two function spaces $\mathcal Q_1$ and $\mathcal Q_2$ consists of $ \sum_{i=1}^k q_{i1}q_{i2}$
for $q_{i1} \in \mathcal Q_1$, $q_{i2} \in \mathcal Q_2$, and any $k \ge 1$.
Then the basis set for such multivariate splines can be written as
\[
\mathcal{I} \bigcup \Big( \bigcup_{k=1}^d \bigcup_{S_k: |S_k|=k}  \mathcal{B}_{S_k} \Big),
\]
where the union is over all $2^d-1$ nonempty subsets $S_k \subset \{1, \ldots, d\}$ and for $S_k = \{j_1,\ldots,j_k\}$,
\[
\mathcal{B}^m_{S_k} =  \big\{ \phi^{(m)}_{\nu_{j_1}, j_1}\cdots\phi^{(m)}_{\nu_{j_k}, j_k}: 2 \le \nu_{j_l} \le n_{j_l}, l=1,\ldots,k \big\}
\]
obtained by taking products of univariate basis functions in $(\mathcal B^m_{j_1}, \ldots, \mathcal B^m_{j_k})$.
As a result, a multivariate spline $g(z_1,\ldots,z_d)$ of cross-order $m$ can be represented as
\begin{equation} \label{eqn:phi-representation}
  g(z_1, \ldots, z_d) = \alpha_0 + \sum_{k = 1}^{d}\sum_{S_k: |S_k|=k} \alpha^\T_{S_k} \Phi^{(m)}_{S_k}
\end{equation}
where $\alpha_{S_k}$ is a column vector of coefficients and $\Phi^{(m)}_{S_k}$ is a column vector of basis functions in $\mathcal{B}^m_{S_k}$, both of dimension $\prod_{l=1}^k (n_{j_l}-1)$
for $S_k = \{j_1, \ldots, j_k\}$.
%For simplicity, the dependency of the basis set $\{\Phi^{(m)}_{S_k}\}$ on $m$ is suppressed unless otherwise stated.
For $m=1$, a multivariate spline of cross-order 1 is piecewise constant in the usual sense.
For $m=2$, a multivariate spline of cross-order 2 is not globally piecewise linear in $d\ge 2$ dimensions,
but said to be piecewise cross-linear: it remains piecewise linear in each coordinate with all others fixed.

For a multivariate spline, the hierarchical TV is in general of a complicated form, depending on linear combinations of the coefficients associated with the basis system $\{\Phi^{(m)}_{S_k}\}$.
We now construct a new basis system $\{ \Psi^{(m)}_{S_k} \}$ by a transformation from $\{\Phi^{(m)}_{S_k} \}$, such that each multivariate spline $g$ in the form (\ref{eqn:phi-representation})
can be represented as
\begin{align}
  g(z_1, \ldots, z_d) = \beta_0 + \sum_{k=1}^d \sum_{S_k: |S_k|=k} \beta^\T_{S_k}\Psi^{(m)}_{S_k} , \label{eqn:psi-representation}
\end{align}
where $\beta_{S_k}$ is a column vector of coefficients. The above representation (\ref{eqn:psi-representation}) will be shown to not only directly yield the
ANOVA decomposition (\ref{eqn:anova-decomp}) with $g_{S_k} = \beta^\T_{S_k}\Psi^{(m)}_{S_k}$, but also
allows a simple expression of the hierarchical TV in terms of the coefficients $\{ \beta_{S_k} \}$.

For $m=1$, define a new basis block for $S_k \subset \{1,\ldots,d\}$ of size $k \ge 1$ as
\begin{align}
& \Psi^{(1)}_{S_k} = \Big\{\prod_{j\in S_k}(1-H_j) \Big\} \Phi^{(1)}_{S_k} ,\label{eqn:psi-def}
\end{align}
where $H_j$ is applied elementwise to a vector of functions.
It can be easily verified that $\Psi^{(1)}_{S_k}$ varies only in $\{z_j: j \in S_k\}$ similarly as $\Psi^{(1)}_{S_k}$,
but satisfies the side condition that $H_j \Psi^{(1)}_{S_k} = 0$ for all $j \in S_k$.
Then for a multivariate spline $g$ of cross-order 1, its ANOVA decomposition (\ref{eqn:anova-decomp}) can be written as (\ref{eqn:psi-representation}) with $g_{S_k} = \beta^\T_{S_k}\Psi^{(1)}_{S_k}$
for some coefficient vector $\beta^\T_{S_k}$.
This can be directly shown by substituting (\ref{eqn:phi-representation}) into (\ref{eqn:anova-fk}):
\begin{eqnarray*}
  g_{S_k} % &=& \prod_{j\in S_k}(1-H_j)\prod_{j\not\in S_k}H_j g \\
   &=& \sum_{l=1}^{d}\sum_{S_l: |S_l|=l}\alpha^\T_{S_l}\prod_{j\in S_k}(1-H_j)\prod_{j\not\in S_k}H_j\Phi^{(1)}_{S_l} \\
   &=& \sum_{l=k}^{d}\sum_{S_l\supset S_k}\alpha^\T_{S_l}\prod_{j\in S_k}(1-H_j)\prod_{j\not\in S_k}H_j\Phi^{(1)}_{S_l}  \\
   &=& \sum_{l=k}^{d} \sum_{S_l\supset S_k} \alpha^\T_{S_l}\Big(\prod_{j\in S_l \backslash S_k}H_j\Phi^{(1)}_{S_l\backslash S_k}\otimes \prod_{j\in S_k}(1-H_j)\Phi^{(1)}_{S_k}\Big) ,
\end{eqnarray*}
The second equality holds because if $S_k\not\subset S_l$, then there exists some $j\in S_k$ but $j\not\in S_l$ and hence $(1-H_j)\Phi^{(1)}_{S_l}=0$ as $\Phi^{(1)}_{S_l}$ is free of $z_j$.
The third equality holds because $\Phi^{(1)}_{S_l} = \Phi^{(1)}_{S_l} \otimes \Phi^{(1)}_{S_l \backslash S_k}$, where $\otimes$ denotes the tensor product of two vectors, but arranged into a vector.
That is, $(a_1, a_2) \otimes (b_1, b_2) = (a_1b_1, a_1b_2, a_2b_1, a_2\otimes b_2)$ for scalars $a_1, b_1$ and vectors $a_2, b_2$.

For $m \ge 2$, we define a new basis system $\{ \Psi^{(m)}_{S_k} \}$ recursively as follows. For $1\le j \le d$, let
\begin{align*}
\psi_{\nu, j } (z_j ) = (1-H_j) \, T_j^{(m)} \phi_{\nu, j}^{(m)}(z_j), \quad \nu=2,\ldots,n_j,
\end{align*}\
where $T_j^{(m)}$ is a linear operator defined such that
\begin{align*}
T_j^{(m)} \phi_{\nu, j}^{(m)} = \left\{
\begin{array}{ll}
(1- z_j H_j D_j) \, D_j^- T_j^{(m-1)} D_j \phi_{\nu, j}^{(m)} , & 3 \le \nu \le n_j, \\
\phi_{2,j}^{(m)} = z_j, & \nu = 2,
\end{array} \right.
\end{align*}
with $T_j^{(1)} = I$ the identity operator. Throughout, $D_j^-$ denotes the integration operator in $z_j$ such that
$ D_j^{-} (z_j^{\nu-1}/(\nu-1)!) = z_j^\nu / \nu!$ for $\nu \ge 1$ and
$ D_j^{-} ( (z_j - t)_+^{m-1}/(m-1)! ) = (z_j - t)^m_+  / m!$ for $m \ge 1$. %Then $D_j D_j^- = I$.
For $m=2$ and 3, the above definition gives
\begin{align*}
T_j^{(2)} \phi^{(2)}_{\nu, j} & = (1-z_j H_j D_j) \phi^{(2)}_{\nu,j}, \quad \nu \ge 3,\\
T_j^{(3)} \phi^{(3)}_{\nu, j} & =
\left\{ \begin{array}{ll}
(1-z_j H_j D_j) \big( \phi^{(3)}_{\nu,j} - \frac{z_j^2}{2} H_j D_j^2 \phi^{(3)}_{\nu,j}\big), & \nu \ge 4, \\
(1-z_j H_j D_j) \phi^{(3)}_{\nu,j}, & \nu = 3.
\end{array} \right.
\end{align*}
A key property from our definition is that by direct calculation,
\begin{align}
D_j T_j^{(m)} \phi_{\nu, j}^{(m)} =
\left\{ \begin{array}{ll}
(1-H_j) T_j^{(m-1)} D_j \phi^{(m)}_{\nu,j}, & \nu \ge 3.  \\
1, & \nu =2.
\end{array} \right. \label{eqn:recursive}
\end{align}
For each nonempty subset $S_k =\{j_1, \ldots, j_k\}$, define a new basis block as
\begin{align*}
& \Psi^{(m)}_{S_k} = \Big\{ \prod_{j \in S_k} (1-H_j) T_j^{(m)} \Big\} \Phi^{(m)}_{S_k} \\
& = \Big( \psi^{(m)}_{\nu_{j_1},j_1}\cdots \psi^{(m)}_{\nu_{j_k},j_k}: 2 \le \nu_{j_l} \le n_j, l=1,\ldots, k \Big),
\end{align*}
where $T^{(m)}_j$ is applied to each product basis in $\Phi^{(m)}_{S_k}$ such that only the univariate basis $\psi_{\nu, j } (z_j )$ is affected with all others fixed.
The new basis block $\Psi^{(m)}_{S_k} $ can also be expressed as
\begin{align}
& \Psi^{(m)}_{S_k} = \Big\{\prod_{j\in S_k}(1-H_j) \Big\} \tilde \Phi^{(m)}_{S_k} , \label{eqn:psi-def2}
\end{align}
where $\tilde \Phi^{(m)}_{S_k} =\prod_{j\in S_k} T_j^{(m)} \Phi^{(m)}_{S_k} $ is an invertible transformation of $ \Phi^{(m)}_{S_k}$.
In fact, $\tilde \Phi^{(m)}_{S_k}$ can be arranged as $C^{(m)}_{S_k} \Phi^{(m)}_{S_k} $ for a lower triangular matrix  $C^{(m)}_{S_k}$ with ones on the diagonal.
Similarly as (\ref{eqn:psi-def}), Eqn (\ref{eqn:psi-def2}) implies that
$\Psi^{(m)}_{S_k} $ varies only in $\{z_j: j \in S_k\}$  and
satisfies the side condition that $H_j \Psi^{(m)}_{S_k} = 0$ for all $j \in S_k$.
Moreover, for a multivariate spline of cross-order $m$, its ANOVA decomposition (\ref{eqn:anova-decomp}) can be written as (\ref{eqn:psi-representation}) with $g_{S_k} = \beta^\T_{S_k}\Psi^{(m)}_{S_k}$.

The following result shows that that the hierarchical TV of a multivariate spline reduces a Lasso penalty on the coefficients based on the new basis system $\{\Psi^{(m)}_{S_k}\}$.
For $S_k = \{j_1, \ldots, j_k\}$, the non-differentiation degree of a basis function $\phi^{(m)}_{\nu_{j_1}, j_1}\cdots\phi^{(m)}_{\nu_{j_k}, j_k}$ in $\Phi^{(m)}_{S_k}$
or the corresponding element in $\tilde\Phi^{(m)}_{S_k}$ or $\Psi^{(m)}_{S_k}$ is
defined as the size of the set $\{j_l: \nu_{j_l} \ge m+1, l=1,\ldots, k\}$, that is, the number of truly truncated power functions among $\phi^{(m)}_{\nu_{j_1},j_l}, \ldots, \phi^{(m)}_{\nu_{j_k}, j_k}$.
In particular, the non-differentiation degree of $\psi^{(m)}_{\nu_{j_1}, j_1}\cdots\psi^{(m)}_{\nu_{j_k}, j_k}$ is 0 if $2 \le \nu_{j_1}, \ldots, \nu_{j_k} \le m$.

\begin{prop}\label{prop:lasso}
Let $m\ge 1$ be a fixed order of differentiation, $\rho_0=0$, and $(\rho_1,\ldots,\rho_d)$ be positive constants.
Suppose that a multivariate spline $g(z)$ is represented in its ANOVA decomposition as (\ref{eqn:psi-representation}).
Then the hierarchical total variation is \vspace{-.15in}
\begin{align}
\mathrm{HTV}_d^{m}\big(g; \rho_1, \ldots, \rho_d \big) = \sum_{k=1}^d \sum_{S_k: |S_k|=k} \| R^{(m)}_{S_k} \beta _{S_k} \|_1,  \label{eqn:HTV-lasso}
\end{align}
where $\|\cdot\|_1$ denotes the $L_1$ norm of a vector, and $R^{(m)}_{S_k}$ is a diagonal matrix such that the diagonal element associated with a coefficient in $\beta_{S_k}$ is $\rho_l$, where $0 \le l \le k$
is the non-differentiation degree of the corresponding basis function in $\Psi^{(m)}_{S_k}$ defined above.
\end{prop}

By Proposition~\ref{prop:lasso}, the hierarchical TV of a multivariate spline with representation (\ref{eqn:psi-representation}) is a scaled Lasso penalty
on the coefficients $\{\beta_{S_k}\}$
associated with the basis set $\{\Psi^{(m)}_{S_k}\}$, except those basis functions which are products of only univariate non-truncated polynomial functions.
For each basis block $\Psi^{(m)}_{S_k}$ with $S_k = \{j_1, \ldots, j_k\}$, there are a total of $(m-1)^k$ coefficients not penalized in $\beta_{S_k}$,
corresponding to the terms in the expansion of $ \prod_{l=1}^k (z_{j_l}+\cdots+z_{j_l}^{m-1}/(m-1)!)$.
For example, in the case of $m=1$, all coefficients in $\beta_{S_k}$ are penalized.
In the case of $m=2$, all coefficients in $\beta_{S_k}$ are penalized except that  associated with the basis $z_{j_1}\cdots z_{j_k}$.

The Lasso representation (\ref{eqn:HTV-lasso}) also implies that the hierarchical TV reduces to $\rho_l$ when applied to
each individual basis function $\{\Psi^{(m)}_{S_k}\}$ whose non-differentiation degree is $l$, including the fact that the hierarchical TV reduces to 0 for those products of non-truncated polynomial functions
as mentioned above. This property provides a conceptually simple characterization of the $\Psi$-basis system.
With $\rho_1=\cdots= \rho_d=1$, each $\Psi$-basis function is a linear combination of $\Phi$-basis functions such that its hierarchical TV is either 0 or 1.

\vspace{-.05in}
\begin{eg} \label{eg:constant}
For bivariate splines of cross-order 1 (piecewise constant), the univariate basis functions (excluding the constant) are before and after transformation \vspace{-.05in}
\begin{align*}
   \phi^{(1)}_{i,1} &= (z_1 - t_{i-1,1})_+^0, \quad \psi^{(1)}_{i,1} = (1- H_1) \phi^{(1)}_{i,1} , \quad i=2,\ldots, n_1,\\
   \phi^{(1)}_{j,1} &= (z_2 - t_{j-1,2})_+^0, \quad \psi^{(1)}_{j,2} = (1- H_2) \phi^{(1)}_{j,2} , \quad j=2,\ldots, n_2 .
\end{align*}
The transformed basis system consists of $\Psi$-basis functions as follows:
\begin{align*}
\Psi^{(1)}_1 & = ( \psi^{(1)}_{i,1}:   i=2,\ldots, n_1 )^\T,   \\
\Psi^{(1)}_2 & = ( \psi^{(1)}_{j,2}:   j=2,\ldots, n_2 )^\T,   \\
\Psi^{(1)}_{1,2} & = ( \psi^{(1)}_{i,1} \psi^{(1)}_{j,2}:  i=2,\ldots, n_1 , j=2,\ldots, n_2 )^\T.
\end{align*}
There are a total of $n_1n_2-1$ basis functions (excluding the constant), in agreement with the grid size minus one.
For a bivariate spline represented as $g=\beta_0 + \sum_{j=1}^2 \beta^\T_j \Psi^{(1)}_j + \beta^\T _{12} \Psi^{(1)}_{12}$, (\ref{eqn:HTV-lasso}) indicates
that $\mathrm{HTV}_2^1 (g) = \rho_1 \sum_{j=1}^2 \| \beta_j \|_1 + \rho_2 \| \beta_{12}\|_1$. In fact, the three terms match those in (\ref{eqn:2d-tv-constant}),
that is, $\TV_1 (H_2 g) = \|\beta_1\|_1$, $\TV_1(H_1 g) = \| \beta_2 \|_1$, and $\TV_2 (g) = \|\beta_{12} \|_1$.
\end{eg}

\begin{figure}
\centering
\subfigure[$\phi$ main effect]{
\includegraphics[scale=0.39]{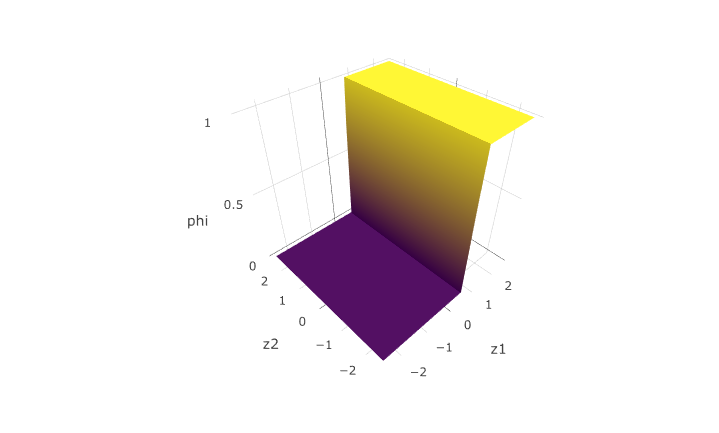}
}
\subfigure[$\phi$ interaction]{
\includegraphics[scale=0.39]{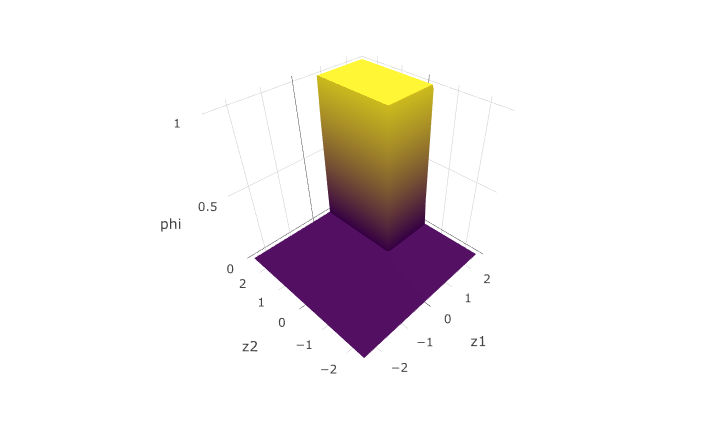}
}
\caption{Piecewise constant $\Phi$-basis functions.}
\label{fig:toy_constant_phi}
\end{figure}
\begin{figure}
\centering
\subfigure[$\psi$ main effect]{
\includegraphics[scale=0.39]{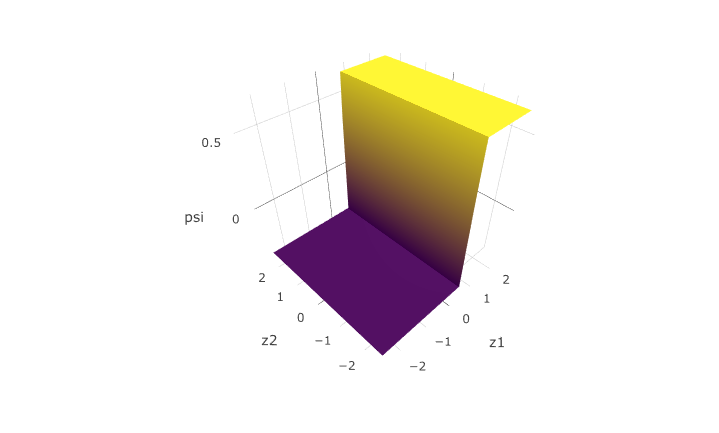}
}
\subfigure[$\psi$ interaction]{
\includegraphics[scale=0.39]{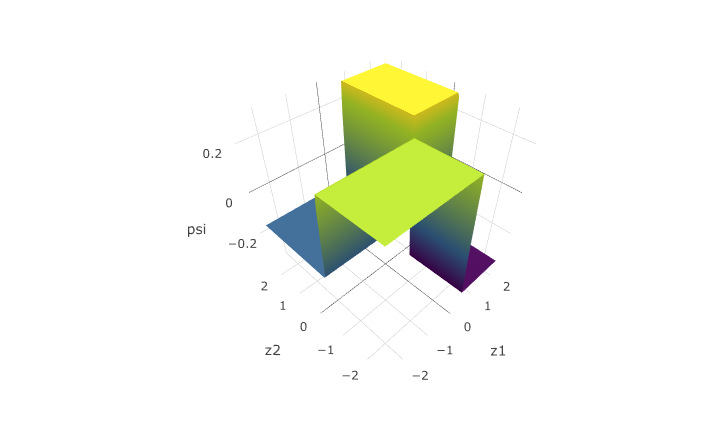}
}
\caption{Piecewise constant $\Psi$-basis functions.}
\label{fig:toy_constant_psi}
\end{figure}

\begin{eg} \label{eg:linear}
For bivariate splines of cross-order 2 (piecewise cross-linear), the univariate basis functions (excluding the constant) are before and after transformation
\begin{align*}
   \phi^{(2)}_{2,1} & = z_1, \quad \psi^{(2)}_{2,1} = (1- H_1) z_1, \\
   \phi^{(2)}_{2,2} & = z_2, \quad \psi^{(2)}_{2,2} = (1- H_2) z_2, \\
   \phi^{(2)}_{i,1} & = (z_1 - t_{i-2,1})_+, \quad \psi^{(2)}_{i,1} = (1-H_1) (\phi^{(2)}_{i,1} - z_1 H_1D_1 \phi^{(2)}_{i,1}) , \quad i=3,\ldots, n_1,\\
   \phi^{(2)}_{j,1} & = (z_2 - t_{j-2,2})_+, \quad \psi^{(2)}_{j,2} = (1-H_2) (\phi^{(2)}_{j,2} - z_2 H_2D_2 \phi^{(2)}_{j,2}), \quad j=3,\ldots, n_2 .
\end{align*}
The transformed basis system consists of $\Psi$-basis functions as follows:
\begin{align*}
\Psi^{(2)}_1 & = ( \psi^{(2)}_{i,1}:   i=2,\ldots, n_1 )^\T,   \\
\Psi^{(2)}_2 & = ( \psi^{(2)}_{j,2}:   j=2,\ldots, n_2 )^\T,   \\
\Psi^{(2)}_{1,2} & = ( \psi^{(2)}_{i,1} \psi^{(2)}_{j,2}:  i=2,\ldots, n_1 , j=2,\ldots, n_2 )^\T.
\end{align*}
%The $\Phi$-basis functions are defined similarly.
Similarly as in Example~\ref{eg:constant}, there are a total of $n_1n_2-1$ basis functions (excluding the constant), in agreement with the grid size minus one.
Suppose that a bivariate spline is represented as $g=\beta_0 + \sum_{j=1}^2 \beta^\T_j \Psi^{(2)}_j + \beta^\T_{12} \Psi^{(2)}_{12}
= \beta_0 + \sum_{i=2}^{n_1} b_{i1} \psi^{(2)}_{i,1} +\sum_{j=2}^{n_2} b_{1j} \psi^{(2)}_{j,2} + \sum_{i=2}^{n_1} \sum_{j=2}^{n_2} \psi^{(2)}_{i,1} \psi^{(2)}_{j,2}$.
where $\beta_1 = (b_{i1}: i=2,\ldots,n_1)^\T$, $\beta_2 = ( b_{1j}: j=2,\ldots,n_2)^\T$, and $\beta_{12} = (b_{ij}: i=2,\ldots,n_1, j=2,\ldots,n_2)^\T$.
Then (\ref{eqn:HTV-lasso}) indicates that
\begin{align*}
\mathrm{HTV}_2^2 (g) = \rho_1 \sum_{i=3}^{n_1} (|b_{i1}| + |b_{i2}|) + \rho_1 \sum_{j=3}^{n_2} (|b_{1j}| + |b_{2j}| ) +  \rho_2 \sum_{i=3}^{n_1} \sum_{j=3}^{n_2} |b_{ij}| .
\end{align*}
The five terms match those in (\ref{eqn:2d-tv-linear}), that is
\begin{align*}
& \TV_1 (H_2 D_1D_2 g) = \sum_{i=3}^{n_1} |b_{i1}|, \quad \TV_1 (H_2 D_1 g) = \sum_{i=3}^{n_1} |b_{i2}|, \\
& \TV_1 (H_1 D_1D_2 g) = \sum_{j=3}^{n_2} |b_{1j}|, \quad \TV_1 (H_1 D_2 g) = \sum_{j=3}^{n_1} |b_{2j}|, \\
& \TV_2 (D_1 D_2 g) = \sum_{i=3}^{n_1} \sum_{j=3}^{n_2} |b_{ij}| .
\end{align*}
These equations can be seen to justify our formulation of the hierarchical TV. If any of the six terms were dropped from (\ref{eqn:2d-tv-linear}), then
$\mathrm{HTV}_2^2 (g) $ would  fail to account for interactions represented by those corresponding basis functions in $\{ \Psi^{(2)}_{S_k} \}$, because
$\mathrm{HTV}_2^2 (g) $ would remain the same with any change of $g$ by adding a linear combination of those basis functions.
\end{eg}

\begin{figure}[H]
\centering
\subfigure[$\phi$ main effect]{
\includegraphics[scale=0.39]{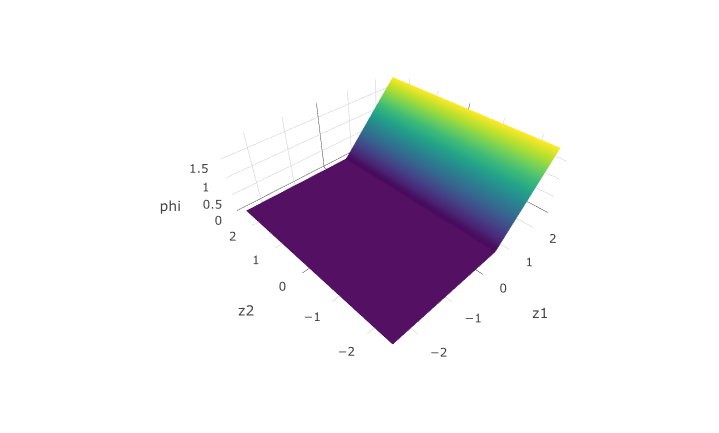}
}
\subfigure[$\phi$ interaction]{
\includegraphics[scale=0.39]{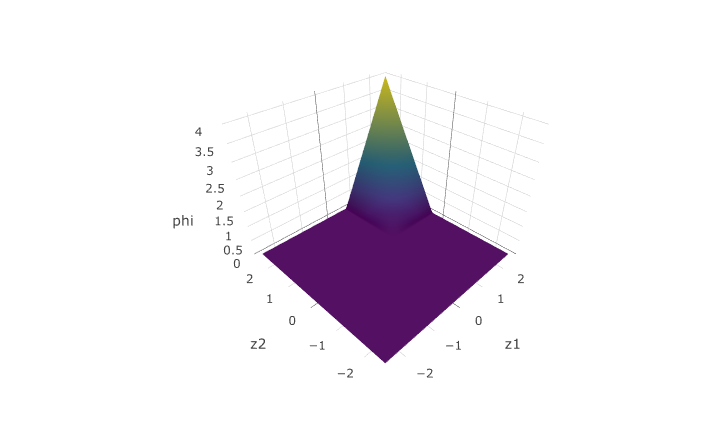}
}
\caption{Piecewise cross-linear $\Phi$-basis functions.}
\label{fig:toy_linear_phi}
\end{figure}

\begin{figure}[H]
\centering
\subfigure[$\psi$ main effect]{
\includegraphics[scale=0.39]{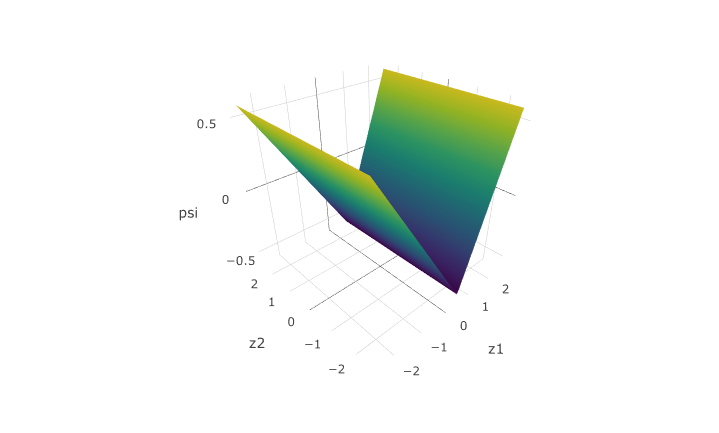}
}
\subfigure[$\psi$ interaction]{
\includegraphics[scale=0.39]{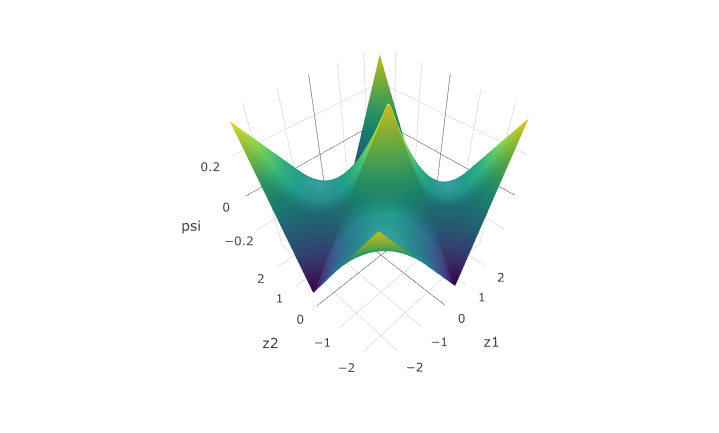}
}
\caption{Piecewise cross-linear $\Psi$-basis functions.}
\label{fig:toy_linear_psi}
\end{figure}

Figures~\ref{fig:toy_constant_phi}--\ref{fig:toy_linear_psi} illustrate the $\Phi$-basis and $\Psi$-basis functions for bivariate splines of order 1 or 2, with the operator $H_j$
specified as the averaging operator $A_j$.


\subsection{Linear ANOVA modeling} \label{sec:linear}

We return to ANOVA modeling for multivariate nonparametric regression.
The data consist of $n$ observations of a response variable $Y_i$ and covariates $X_i = (X_{i1} ,\ldots, X_{ip})$.
The ANOVA representation (\ref{eqn:f-anova}) for the mean function $f$ in (\ref{eqn:model}) can be obtained from (\ref{eqn:anova-decomp}) with $g=f$
by truncating all higher than $K$-way interactions, that is, setting $f_{S_k} =0$ for all $k > K$.

For estimation, we restrict $f$ to be a multivariate spline of cross-order $m \ge 1$, and
take the functional semi-norm $\|\cdot\|_F$ to be the hierarchical TV of the corresponding order $m$. The marginal knots
$z_{1,j} < \cdots < z_{n_j,j}$ can be specified as quantiles of the data points $\{X_{1j}, \ldots, X_{nj}\}$.
Then the ANOVA decomposition of $f$ can be expressed as (\ref{eqn:psi-representation}) in terms of the $\Psi$-basis functions,
with $f_{S_k} = \Psi^{(m)}_{S _k}\beta^\T_{S _k}$ for $1 \le k \le K$.
By Proposition~\ref{prop:lasso}, $\mathrm{HTV}_k^m ( f_{S_k} ) = \| R^{(m)}_{S _k}\beta_{S _k}\|_1$. Hence
the doubly penalized loss function (\ref{eqn:doubly-pen2}) becomes \vspace{-.1in}
\begin{equation}\label{eqn:transformed-lasso}
    \frac{1}{2}\big\|Y -\beta_0 - \sum_{k=1}^K\sum_{S _k: |S _k|=k} \Psi^\dag_{S _k}\beta_{S _k}\big\|_n^2 +
    \sum_{k=1}^K\sum_{S _k:|S _k|=k}\Big\{\big\| R_{S _k}\beta_{S _k}\big\|_1 + \lambda_k \big\|\Psi^\dag_{S _k}\beta_{S _k}\big\|_n\Big\},
\end{equation}
where $\Psi^\dag_{S_k}$ is the data matrix formed from the basis block $\Psi^{(m)}_{S_k}$, i.e., the $i$th row of $\Psi^\dag_{S_k}$ is the transpose of $\Psi^{(m)}_{S_k}(X_i)$ for $i=1,\ldots,n$,
and $R_{S _k}$ is $R^{(m)}_{S_k}$ in Proposition~\ref{prop:lasso}. For simplicity, the dependency on $m$ is suppressed unless otherwise stated.

To avoid interference between $\beta_0$ and $\{ \beta_{S_K} \}$, we replace in (\ref{eqn:transformed-lasso}) $\beta_0$ by $\bar Y$ and $\Psi^\dag_{S _k}$ by the empirically centered version
$\tilde\Psi^\dag_{S _k} = \Psi^\dag_{S _k} - \bar \Psi^\dag_{S_k}$,
where $\bar Y = n^{-1} \sum_{i=1}^n Y_i$ and $\bar \Psi^\dag_{S_k} = n^{-1} \sum_{i=1}^n \Psi^\dag_{S_k}$. This leads to the doubly penalized loss function
\begin{equation}\label{eqn:transformed-lasso2}
    \frac{1}{2}\big\|Y - \bar Y- \sum_{k=1}^K\sum_{S _k: |S _k|=k} \tilde \Psi^\dag_{S _k}\beta_{S _k}\big\|_n^2 +
    \sum_{k=1}^K\sum_{S _k:|S _k|=k}\Big\{\big\| R_{S _k}\beta_{S _k}\big\|_1 + \lambda_k \big\|\tilde \Psi^\dag_{S _k}\beta_{S _k}\big\|_n\Big\} .
\end{equation}
Formally, (\ref{eqn:transformed-lasso2}) can be derived from (\ref{eqn:transformed-lasso}) as follows.
Suppose that $\Psi^\dag_{S _k}\beta_{S _k}$ is modified in (\ref{eqn:transformed-lasso}) as $\beta_{S_k,0} + \Psi^\dag_{S _k}\beta_{S _k}$ with an intercept $\beta_{S_k,0}$ for each block $S_k$. Then
minimization of the modified loss function over $\beta_0$ and $\{\beta_{S_k,0}\}$ yields the loss function (\ref{eqn:transformed-lasso2}).

To numerically minimize (\ref{eqn:transformed-lasso2}) over $\{ \beta_{S_k} \}$, we exploit a backfitting algorithm, called block descent and thresholding (BDT), in \cite{yang2018backfitting},
which is presented as Algorithm~\ref{alg:bdt-fanova}. The algorithm was originally developed to minimize a similar doubly penalized loss function as (\ref{eqn:transformed-lasso2}) for
high-dimensional additive modeling with $K=1$.
In general, a backfitting algorithm involves iteratively updating each coefficient vector $\beta_{S_k}$ while fixing all the others. For a block $S_k \subset \{1,\ldots,p\}$,
the sub-problem is
\begin{equation}\label{eqn:subproblem}
  \min_{\beta_{S _k}} \; \frac{1}{2}\big\|r_{S _k}-\tilde{\Psi}^\dag_{S _k}\beta_{S _k}\big\|_n^2 + \big\|R_{S _k}\beta_{S _k}\big\|_1 + \lambda_k \big\|\tilde{\Psi}^\dag_{S _k}\beta_{S _k}\big\|_n
\end{equation}
where $r_{S _k} = Y - \bar{Y} - (\hat{f} - \tilde{\Psi}^\dag_{S _k}\hat{\beta}_{S _k})$, $\hat{f} = \sum_{l=1}^K\sum_{S _l: |S _l|=l} \tilde \Psi^\dag_{S _l} \hat\beta_{S _l}$,
and $\hat \beta_{S_l}$'s are the current estimates.
The BDT algorithm is then based on the following result \citep{yang2018backfitting}: if $\tilde \beta_{S_k}$ is a solution to the Lasso problem
\begin{equation}\label{eqn:subproblem-lasso}
  \min_{\beta_{S _k}} \, \frac{1}{2}\big\|r_{S _k}-\tilde{\Psi}^\dag _{S _k}\beta_{S _k}\big\|_n^2 + \big\|R_{S _k}\beta_{S _k}\big\|_1 ,
\end{equation}
then a solution to the double-penalty problem (\ref{eqn:subproblem}) is $\hat \beta_{S_k} = (1 - \lambda_k / \| \tilde{\Psi}^\dag _{S _k} \tilde \beta_{S _k} \|_n)_+ \tilde \beta_{S_k}$, that is,
$\hat \beta_{S_k}$ is determined from $\tilde \beta_{S_k}$ by a vector version of soft thresholding.
In addition, screening rules from  \cite{yang2018backfitting} can also be employed to identify a zero solution $\hat\beta_{S_k}$ to problem (\ref{eqn:subproblem}) directly, without solving the Lasso problem (\ref{eqn:subproblem-lasso}).

The above characterization of a solution to (\ref{eqn:subproblem}) not only underlies the derivation of the BDT algorithm, but also makes explicit the distinct effects
of using the two penalties. Use of the Lasso penalty $\|R_{S _k}\beta_{S _k}\|_1$, obtained from the hierarchical TV, can induce a sparse solution of
 $\tilde \beta_{S_k}$ to problem (\ref{eqn:subproblem-lasso}), corresponding to a sparse selection of basis functions (or multi-dimensional knots) within the block $\tilde{\Psi}^\dag_{S _k}$.
For $S_k = \{j_1, \ldots, j_k\}$, the total number of basis functions within $\tilde{\Psi}^\dag_{S _k}$ is $\prod_{l=1}^k (n_{j_l}-1)$ and often very large.
Use of the empirical-norm penalty $\|\tilde{\Psi}^\dag_{S _k}\beta_{S _k}\|_n$ can result in a zero solution $\hat \beta_{S_k}$ to (\ref{eqn:subproblem}) via thresholding the Lasso solution $\tilde \beta _{S_k}$,
and hence achieve a sparse selection of component functions $f_{S_k} = \tilde{\Psi}^\dag_{S _k}\beta_{S _k}$.
The total number of component functions (or blocks) is ${p \choose 1} + \cdots + {p \choose K}$ and can be very large for $K \ge 2$.

\begin{algorithm} %[H]
\caption{Block Descent and Thresholding (BDT) algorithm}\label{alg:bdt-fanova}
\begin{algorithmic}[1]
\STATE  {\bf Initialize:} Compute $\tilde{\Psi}_{S _k}$ and set $\hat{\beta}_{S _k} = 0$ for all $S _k$ and $1\leq k\leq K$.
\FOR {$k=1, \dots, K$ and $S _k \subset \{1, \ldots, p\}$ with $|S _k|=k$}
\IF {any screening condition is satisfied \citep{yang2018backfitting}}
\STATE Return $\hat{\beta}_{S _k}$.
\ELSE
\STATE Update the residual: $r_{S _k} = Y - \bar{Y} - (\hat{f} -\tilde{\Psi}^\dag_{S _k}\hat{\beta}_{S _k})$.
\STATE Compute a solution $\tilde \beta_{S _k}$ to problem (\ref{eqn:subproblem-lasso}).
\STATE Threshold the solution: $\hat {\beta}_{S _k} = \Big(1-\frac{\lambda_k}{\|\tilde{\Psi}^\dag_{S _k} \tilde {\beta}_{S _k}\|_n}\Big)_+ \tilde {\beta}_{S _k}$.
\ENDIF
    %\item 1. If screening condition (Section~\ref{sec:screen}) is satisfied:
    %\item 2. else
   % \item 3. Compute a solution $\hat\beta_j$ to problem (\ref{eqn:sub-problem2}) using a descent algorithm.
    %\item 4. Threshold the solution: $\tilde{\beta}_j = \Big(1-\frac{\lambda}{\|\Psi\tilde{\beta}_j\|_n}\Big)_+\hat{\beta}_j$.
\ENDFOR
\STATE Repeat line 2-10 until convergence of the objective (\ref{eqn:transformed-lasso2}).
\end{algorithmic}
\end{algorithm}

To solve Lasso problem (\ref{eqn:subproblem-lasso}), it is possible to use a variety of numerical methods including coordinate descent \citep{friedman2010regularization,wu2008coordinate},
gradient descent-related methods \citep{beck2009fast,kim2007interior}, and active-set descent \citep{osborne2000new,yang2018backfitting}.
We currently employ an active-descent method, which was shown to enjoy several advantages for solving Lasso sub-problems during backfitting \citep{yang2018backfitting}.
First, active-set descent finds an exact solution after a finite number of iterations, in contrast with other methods such as coordinate descent.
More accurate within-block estimates $\hat \beta_{S_k}$ can result in fewer backfitting cycles to
achieve convergence by a pre-specified criterion. Second, the computational cost of active-set descent is often reasonable in sparse settings.
In fact, the method allows Lasso problem (\ref{eqn:subproblem-lasso}) to be solved in one or a few iterations, if the estimate $\tilde \beta_{S_k}$ from the previous backfitting cycle,
when used as an initial value, is close to the desired solution.
Third, the active-descent method is tuning free. Gradient descent related methods involve tuning step sizes to achieve satisfactory convergence.
It can be cumbersome to select such tuning parameters for all Lasso sub-problems in backfitting.

We use another active-set technique to speed up the backfitting algorithm with a large number of blocks in sparse settings, similarly as in
\cite{krishnapuram2005sparse} %\cite{meier2008group},
and \cite{friedman2010regularization}.
After completing a full cycle through all the blocks,
we iterate on only the \emph{active} blocks $S_k$ with nonzero estimates $\hat\beta_{S_k}$ till convergence. If another full cycle does not change the active blocks, convergence is obtained;
otherwise the process is repeated on the new set of active blocks.
This technique is a scheme for organizing backfitting cycles, and should be distinguished from the active-set optimization method above.

\subsection{Logistic ANOVA modeling}\label{sec:logistic}

As an extension to handle binary responses $Y_i$, consider a logistic functional ANOVA model:
\begin{align*}
P(Y_i = 1 | X_i ) = \mathrm{expit} ( f(X_i) ),
\end{align*}
where $\mathrm{expit}(c) = \{1 + \exp(-c)\}^{-1}$ and $f(x) = f(x_1,\ldots,x_p)$ is a ``linear" predictor with ANOVA representation (\ref{eqn:f-anova}).
For estimation similarly as in Section~\ref{sec:linear}, we restrict $f$ to be a multivariate spline of cross-order $m \ge 1$, and
take the functional semi-norm $\|\cdot\|_F$ to be the hierarchical TV of the corresponding order $m$.
A doubly penalized loss function based on the log-likelihood for logistic modeling, similar to (\ref{eqn:transformed-lasso2}), is
\begin{equation} \label{eqn:logistic-loss}
  \frac{1}{n}\sum_{i=1}^n\ell\big(Y_i, \mu+f_i \big) + \sum_{k=1}^K\sum_{S _k:|S _k|=k}\Big\{\big\|D_{S _k}\beta_{S _k}\big\|_1 + \lambda_k \big\|\tilde{\Psi}^\dag_{S _k}\beta_{S _k}\big\|_n\Big\},
\end{equation}
where $\ell (Y_i,\mu+f_i) = \log\{1 + \exp(\mu+f_i)\} - y_i(\mu+ f_i)$ and, by abuse of notation, $f = \sum_{k=1}^K\sum_{S _k:|S _k|=k} \tilde{\Psi}^\dag_{S _k}\beta_{S _k}$, a column vector
with elements $(f_1,\ldots,f_n)$.
In contrast with (\ref{eqn:transformed-lasso2}) for linear modeling, the intercept $\mu$ cannot be directly estimated as $\bar Y$.

For backfitting, the sub-problem corresponding to a block $S_k \subset \{1,\ldots,p\}$ is
\begin{equation} \label{eqn:logistic-subproblem}
   \min_{\mu,\beta_{S _k}} \; \frac{1}{n}\sum_{i=1}^n\ell\big(y_i, \mu+\hat{f}-\tilde{\Psi}^\dag_{S _k}(\hat{\beta}_{S _k}-\beta_{S _k})\big) + \big\|R_{S _k}\beta_{S _k}\big\|_1 + \lambda_k \big\|\tilde{\Psi}^\dag_{S _k}\beta_{S _k}\big\|_n ,
\end{equation}
where $\hat{f} = \sum_{l=1}^K\sum_{S _l:|S _l|=l}\tilde{\Psi}^\dag_{S _l}\hat{\beta}_{S _l}$, and $\hat{\beta}_{S _l}$'s are the current estimates.
Similarity as in \cite{friedman2010regularization}, we form a quadratic approximation to the negative log-likelihood term via a Taylor expansion about the previous estimates $\hat{\mu}$ and
$\hat{\beta}_{S _k}$, and obtain the following weighted least squares problem:
\begin{equation} \label{eqn:logistic-subproblem-quadratic}
  \min_{\mu, \beta_{S _k}} \frac{1}{2n}\sum_{i=1}^n w_i\big(\eta_{S _k, i} - \mu -\tilde{\Psi}^\dag_{S _k}\beta_{S _k}\big)^2 + \big\|R_{S _k}\beta_{S _k}\big\|_1 + \lambda_k \big\|\tilde{\Psi}^\dag_{S _k}\beta_{S _k}\big\|_n
\end{equation}
where $\eta_{S _k, i} = \hat{\mu} + \tilde{\Psi}^\dag_{S _k}\hat{\beta}_{S _k} + w_i^{-1}(y_i - \hat{p}_i)$, $\hat{p}_i = \mathrm{expit}(\hat{\mu} + \hat{f}_i)$, and $w_i = \hat{p}_i(1-\hat{p}_i)$.
Application of active-set descent to solve (\ref{eqn:logistic-subproblem-quadratic}) would involve
computing a new Cholesky decomposition of the Gram matrix with weights $(w_1,\ldots,w_n)$ from the active columns in $\tilde{\Psi}^\dag_{S _k}$.
To reduce computational cost, we solve (\ref{eqn:logistic-subproblem-quadratic}) with each $w_i$ replaced by its upper bound $1/4$.
The resulting solution $(\hat\mu, \hat \beta_{S_k})$ remains a descent update, in decreasing the objective value in (\ref{eqn:logistic-subproblem}),
by the quadratic lower bound principle \citep{bohning1988monotonicity,wu2008coordinate}. We summarize
the backfitting algorithm for logistic ANOVA modeling as Algorithm~\ref{alg:bdt-logit}.

\begin{algorithm}[H]
\caption{Block Descent and Thresholding algorithm for logistic modeling (BDT-Logit)}\label{alg:bdt-logit}
\begin{algorithmic}[1]
\STATE  {\bf Initialize:} Set $\hat{\mu}=0$ and $\hat{\beta}_{S _k} = 0$ for all $S _k$ and $1\leq k\leq K$. Set $w_0=1/4$.
\FOR {$k=1, \dots, K$ and $S _k \in \{1, \ldots, p\}$ with $|S _k|=k$}
\STATE    Compute $\hat{p}= \mbox{expit}(\hat{\mu} + \sum_{k=1}^K\sum_{S _k:|S _k|=k}\tilde{\Psi}_{S _k} \hat{\beta}_{S _k})$ and $\eta_{S _k} = \hat{\mu}+\tilde{\Psi}_{S _k}\hat{\beta}_{S _k} + w_0^{-1} (y-\hat{p})$.
\STATE    Update $\hat{\mu} =\bar {\eta}_{S _k}$, the sample average of $\eta_{S _k}$.
\STATE    Update $\hat{\beta}_{S _k}$ as a solution to (using Algorithm~\ref{alg:bdt-fanova}, line 3-9)
    \[ \min_{\beta_{S _k}} \left\{ \frac{w_0}{2}\|\eta_{S _k}-\bar \eta_{S _k}- \tilde\Psi_{S _k}\beta_{S _k}\|_n^2+\|D\beta_{S _k}\|_1 + \lambda \| \tilde\Psi_{S _k}\beta_{S _k}\|_n \right\} . \]
\ENDFOR
\STATE Repeat line 2-7 until convergence of the objective (\ref{eqn:logistic-loss}).
\end{algorithmic}
\end{algorithm}

\section{Numerical experiments}\label{sec:experiment}

We perform numerical experiments to evaluate the empirical performance of the proposed method, doubly penalized ANOVA modeling (dPAM) using hierarchical total variations.
R codes for implementing our method are provided in the Supplementary Material.

Both linear modeling (``regression") and logistic modeling (``classification") are studied.
The performance is measured in terms of mean integrated squared errors (MISEs) and mean square errors (MSEs) for regression, and logistic loss, accuracy, and AUC for binary classification.
We report results only from dPAM with two-way interactions (i.e., $K=2$), which, as will be shown, achieves competitive or superior performance compared with existing methods, possibly with higher than two-way interactions.
For simplicity, the tuning parameters are set such that $\rho_1=\rho_2=\rho$ and $\lambda_1=\lambda_2=\lambda$ for the total-variation and empirical-norm penalties.
The marginal knots for each covariate used are 11 quantiles by 10\% in regression (simulation and real data) and classification (simulation), and 6 quantiles by 20\% in classification (real data).
For fixed-point TV, the fixed point is the minimum corner of the knot grid.
See Section~\ref{sec:2d-smooth} in the Supplement for a simple example on comparison of fixed-point and averaging TVs.

We compare our method with several existing methods including multivariate adaptive regression splines (MARS) \citep{friedman1991multivariate},
tree boosting \citep{freund1997decision,trevor2009elements}, random forest (RF) \citep{breiman2001random}, and
the component selection and smoothing operator (COSSO) \citep{lin2006component}.
We use the R package \texttt{mda} \citep{mda} for MARS,  \texttt{gbm} \citep{gbm} for tree boosting,
\texttt{randomForest} \citep{randomforest} for random forest.
The released package \texttt{cosso} \citep{cosso} for COSSO currently handles only univariate additive modeling. Hence the results
for regression with two-way interactions by COSSO are directly taken from \cite{lin2006component}.

\subsection{Linear ANOVA modeling}

\subsubsection{Simulation study} \label{sec:linear-simulation}

As in \cite{lin2006component}, consider four functions on interval $[0, 1]$, \vspace{-.1in}
\begin{eqnarray*}
 && g_1(z) = z , \quad g_2(z) = (2z-1)^2  , \quad g_3(z) = \frac{\sin(2\pi z)}{2- \sin(2\pi z)}  ,\\
 && g_4(z) = 0.1\sin(2\pi z) + 0.2 \cos(2\pi z) + 0.3 \sin^2(2\pi z) + 0.4 \cos^3(2\pi z) + 0.5 \sin^3(2\pi z) .
\end{eqnarray*}
Define a mean function with two-way interactions as \vspace{-.1in}
\begin{eqnarray*}
f(x) &=& g_1(x_1) + g_2(x_2) + g_3(x_3) + g_4(x_4) + g_1(x_3x_4) + g_2(\frac{x_1+x_3}{2}) + g_3(x_1x_2) .
\end{eqnarray*}
For $i=1,\ldots,n$, the data $Y_i$ and $X_i=(X_{i1}, \ldots, X_{i4})$ are generated such that $Y_i = f(X_i) + \epsilon_i$, where
$X_{ij} \sim \mathrm{Uniform}\,[0, 1]$ independently for $j=1,\ldots,4$, and $\epsilon_i \sim \mathrm{N}(0, 0.2546^2)$, giving a signal-to-noise ratio of 3:1 \citep{lin2006component}.
In addition, six spurious covariates $(X_{i5}, \ldots, X_{i,10})$ are generated from $\mathrm{Uniform}\,[0, 1]$, independently of $(Y_i,X_{i1},\ldots,X_{i4})$ and added to the covariate vector $X_i$.
Hence $X_i=(X_{i1}, \ldots, X_{i,10})$ is of dimension $p=10$.
There are a total of $55$ basis blocks for our method; $45$ of them are interactions.

For $n=100, 200, 400$, we generate training and validation sets each with $n$ observations.
The tuning parameters $(\lambda, \rho)$ are selected to minimize the mean squared error (MSE) on the validation set.
This scheme is used to mimic cross validation, but with lower computational cost to allow repeated runs, as previously in \cite{petersen2016fused} and \cite{yang2018backfitting}.
Similarly, the hyper-parameters \texttt{shrinkage} and \texttt{n.trees} are tuned for \texttt{gbm}, and \texttt{mtry} is tuned for \texttt{randomForest}.
The function \texttt{mars} is applied without tuning in \texttt{mda}.

For each estimator $\hat f(x)$ obtained, the mean integrated squared error (MISE) is evaluated by Monte Carlo integration using fixed $N=10,000$ test data points $\xi_i$, that is,
\begin{equation*}
\mathrm{MISE} = \frac{1}{N} \sum_{i=1}^N (f( \xi_i ) - \hat f(\xi_i))^2 .
\end{equation*}
The simulation is then repeated for $100$ times and the means and standard errors of MISEs are summarized from various methods in Table~\ref{tab:simulated-mise}.

\begin{table}[t]
  \centering
  \caption{MISEs from linear ANOVA simulation}\label{tab:simulated-mise}
  \begin{tabular}{>{}c*{3}{c}}\hline
   & \multirow{1}{*}{\bfseries $n=100$} & \multirow{1}{*}{\bfseries $n=200$} & \multirow{1}{*}{\bfseries $n=400$}\\
\hline
  $\mathrm{ATV}$, $m=1$ & 0.194 (0.003) & 0.118 (0.002) & 0.075 (0.001) \\
  $\mathrm{FTV}$, $m=1$ & 0.231 (0.003) & 0.155 (0.001) & 0.096 (0.001) \\
  $\mathrm{ATV}$, $m=2$ & {\bf 0.118 (0.002)} & {\bf 0.055 (0.001)} & {\bf 0.026 (0.001)} \\
  $\mathrm{FTV}$, $m=2$ & 0.125 (0.003) & 0.062 (0.001) & 0.032 (0.001)  \\
  COSSO (GCV)$^*$           & 0.358 (0.009) & 0.100 (0.003) & 0.045 (0.001)  \\
  COSSO (5CV)$^*$           & 0.378 (0.005) & 0.094 (0.004) & 0.043 (0.001) \\
  MARS, \texttt{degree} $=1$ & 0.247 (0.005) & 0.184 (0.002) & 0.161 (0.001) \\
  MARS, \texttt{degree} $=2$ & 0.281 (0.020) & 0.117 (0.004) & 0.083 (0.001) \\
  MARS, \texttt{degree} $=3$ & 0.300 (0.021)$^*$ & 0.122 (0.004)&  0.084( 0.001) \\
  GBM, \texttt{depth} $=1$ & 0.283 (0.003) & 0.206 (0.002) & 0.171 (0.001) \\
  GBM, \texttt{depth} $=2$ & 0.229 (0.003) & 0.118 (0.001) & 0.071 (0.001) \\
  GBM, \texttt{depth} $=3$ & 0.227 (0.003) & 0.113 (0.001) & 0.066 (0.001) \\
% RF, \texttt{ntree} $=50$ & 0.293 (0.004) & 0.212 (0.002) & 0.159 (0.001) \\
  RF, \texttt{ntree} $=100$ & 0.291 (0.004) & 0.209 (0.002) & 0.157 (0.001) \\
  RF, \texttt{ntree} $=200$ & 0.289 (0.004) & 0.208 (0.002) & 0.155 (0.001) \\
  RF, \texttt{ntree} $=400$ & 0.288 (0.004) & 0.206 (0.002) & 0.154 (0.001) \\
\hline
  \end{tabular}\\[.1in]
\parbox{1\textwidth}{\small Note: For COSSO, the results are taken from \cite{lin2006component} and may not be comparable with other results, because
  the hyper-parameters are selected by GCV or 5-fold cross validation (5CV).
  For our method, ATV or FTV indicates averaging or fixed-point hierarchical TV used. For MARS or GBM, \texttt{degree} or \texttt{depth} is the level of interactions.
  For RF, \texttt{ntree} is the number of trees to grow. For $n=100$, the result for MARS, \texttt{degree} $=3$ is reported after removing one extremely large MISE.}
\end{table}

From Table~\ref{tab:simulated-mise}, the two versions of our method using piecewise cross-linear splines ($m=2$) and either
averaging TV (ATV) or fixed-point TV (FTV) achieve the best two performances, sometimes with substantial margins of improvement,
among all methods studied. Use of the averaging TV performs better than the fixed-point.
Our method using piecewise constant splines ($m=1$) yields an estimator $\hat f(x)$ in a comparable form as that from tree boosting with interaction depth 2.
The performance of our piecewise constant method using averaging TV is similar to that of the latter, except noticeably better for sample size $n=100$.

\begin{figure}[H]
  \centering
  \includegraphics[width=5in, height=3.5in]{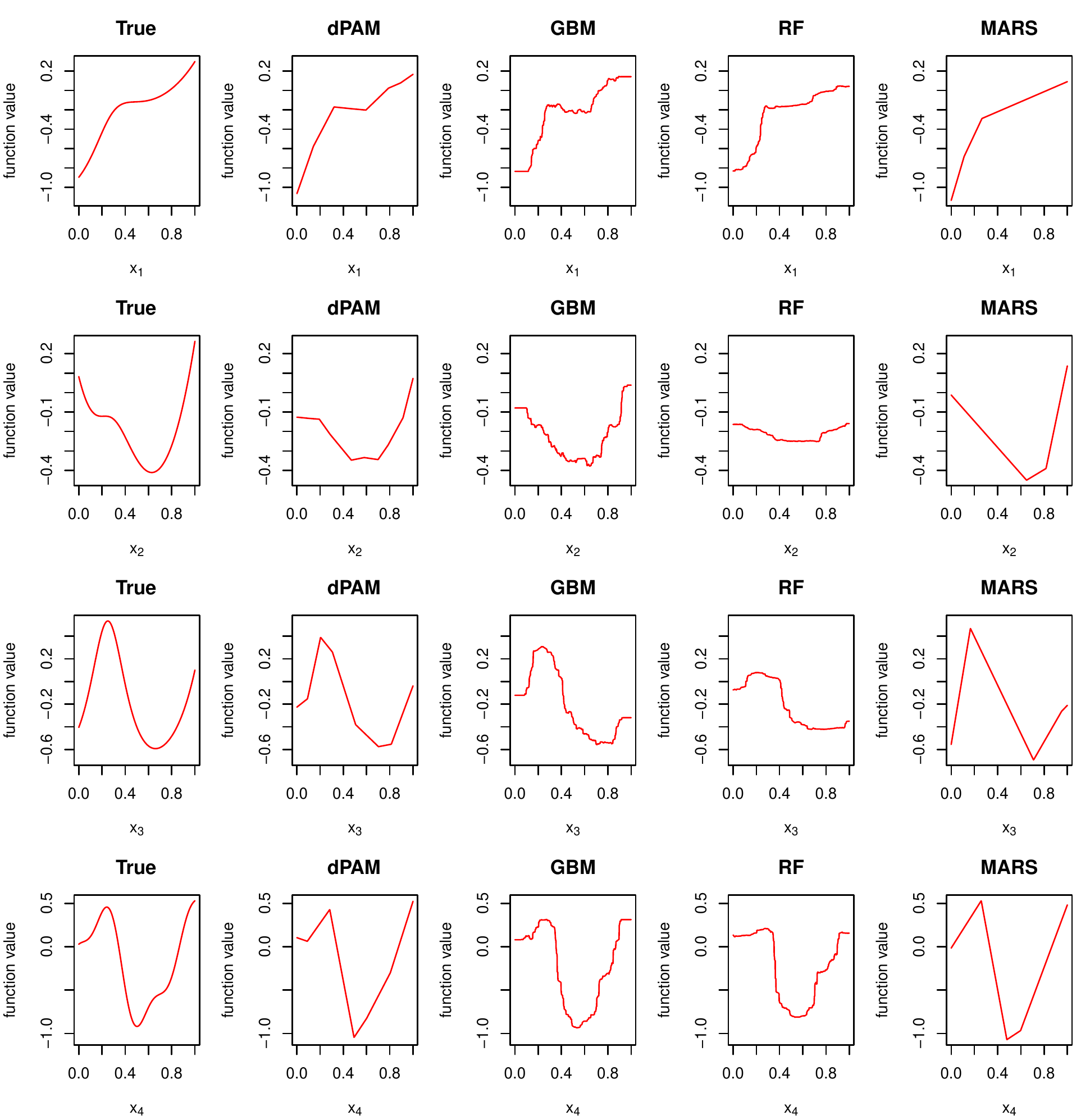}\\
  \caption{$1$-dimensional partial dependence plots for $p=10$ and $n=200$.}\label{fig:1d-partial-dependence-plot-n-200}
\end{figure}

\begin{figure}[H]
  \centering
  \includegraphics[width=5in, height=3in]{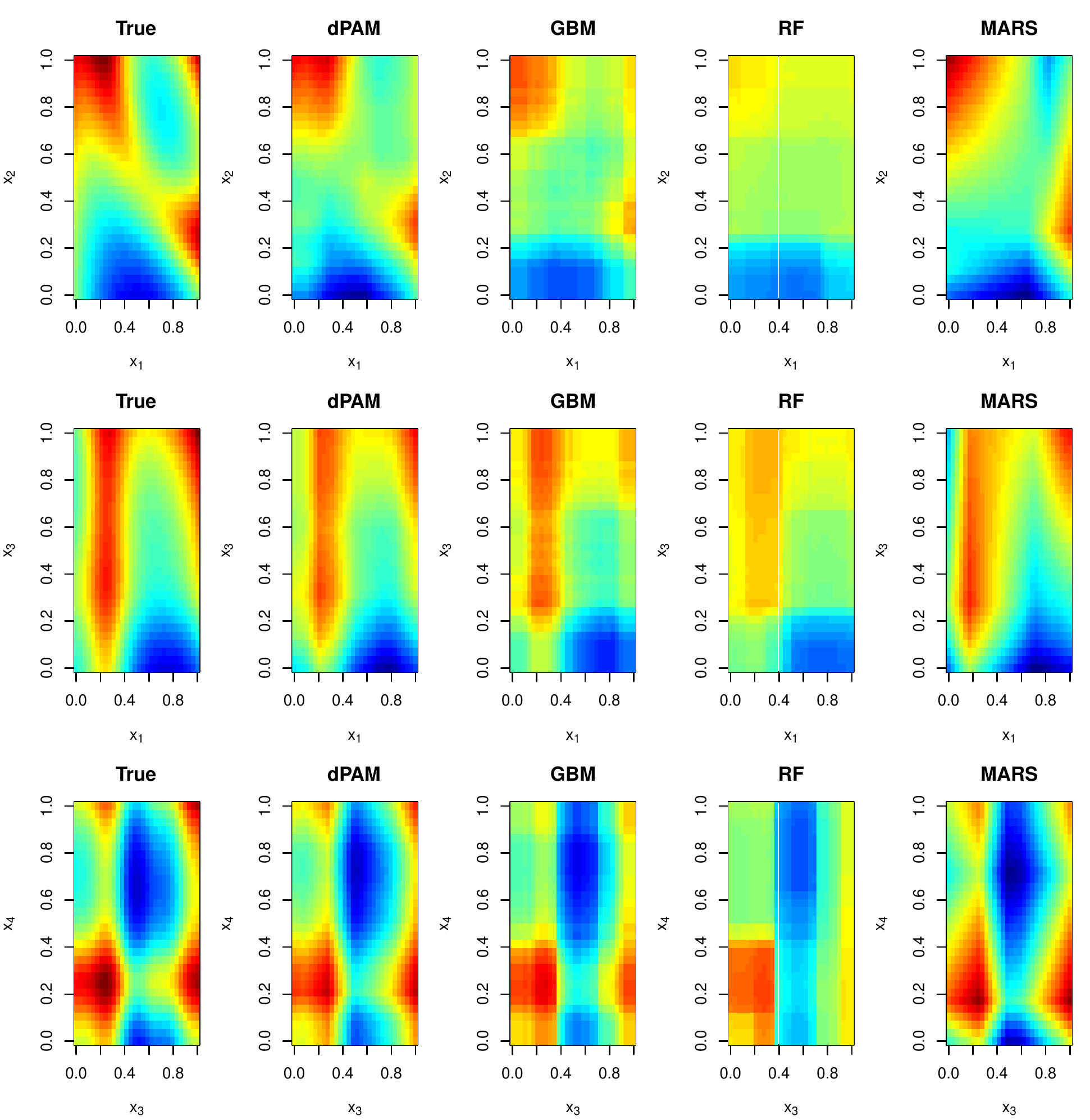}\\
  \caption{$2$-dimensional partial dependence plots for $p=10$ and $n=200$.}\label{fig:2d-partial-dependence-plot-n-200}
\end{figure}

For interpretation, partial dependence plots can be used to visualize the dependency of an approximation $\hat f(x)$ on its arguments,
as in Hastie et al.~(2009), Section 10.13.2.
Let $X_{\mathcal{S}}$ be a subvector of variables of interest, indexed by $\mathcal S \subset \{1,\ldots,p\}$. Let $\mathcal{C}$ be the complement of $\mathcal S$.
A partial dependence function can be estimated by
\begin{equation*}
  \bar{f}_{S }(X_{\mathcal{S}}) = \frac{1}{n}\sum_{i=1}^n \hat{f}(X_{\mathcal{S}}, x_{i\mathcal{C}})
\end{equation*}
where $\{x_{1\mathcal{C}}, \ldots, x_{n\mathcal{C}}\}$ are the values of $X_{\mathcal{C}}$ occurring in the training data.
For sample size $n=200$, Figures~\ref{fig:1d-partial-dependence-plot-n-200} and \ref{fig:2d-partial-dependence-plot-n-200} show
the 1- and 2-dimensional partial dependence plots  for our method with ATV and $m=2$, MARS with \texttt{degree}=2,
GBM with \texttt{depth}=2, and RF with \texttt{ntree}=400. See the Supplement for the corresponding figures with sample sizes $n=100$ or 400.
Among all the methods studied, our method yields the most accurate approximation to the ground truth, which confirms the comparison in Table~\ref{tab:simulated-mise}.
Moreover, the partial dependence functions from our method are simpler, with a much fewer number of knots apparently in 1-dimensional plots, than those from
other methods archiving similar patterns.

\subsubsection{Empirical examples}

We evaluate our method on three popular real datasets. Boston housing and Ozone data are available from R library \texttt{mlbench} \citep{mlbench}. For Boston housing data, the responses are housing values in Boston in 1970. There are $12$ input variables and $506$ observations. For the Ozone data, the responses are the daily maximum one-hour-average ozone readings in Los Angeles in 1976. There are $8$ input variables and $330$ observations, after removal of missing data. The Tecator data is available from the dataset archive of StatLib~\url{http://lib.stat.cmu.edu/datasets/tecator}.
The response variable is the fat content of a meat sample. As suggested, we use the first $13$ principal components of the absorbance spectrum as input variables. The total sample size is $215$.

We apply various methods on these datasets, including doubly penalized additive modeling with linear splines (dPAM1, linear) from \cite{yang2018backfitting}.
For each method, the prediction mean squared errors $\mathrm{E}\{(Y-\hat{f}(X))^2\}$ is estimated by $10$-fold cross-validation, as in \cite{lin2006component}.
For each 9:1 split of training and test sets, the tuning parameters are selected by $5$-fold cross-validation within the training set and the estimate obtained is then evaluated on the test set.
The $10$-fold cross-validation is repeated for $5$ times. The same random splits are used for all methods.
%The same randomization was used to tabulate the performances of the reference methods.
The means and standard errors of estimated MSEs are summarized in Table~\ref{tab:real-mse}. The results for COSSO are taken from \cite{lin2006component}.

From Table~\ref{tab:real-mse}, the two versions of our method using piecewise cross-linear splines are consistently among the three methods with the lowest MSEs.
There seems to be no definite comparison between use of averaging or fixed-point TV: the former performs better on Tecator data, but worse on Boston and Ozone data.
It is reassuring that our ANOVA modeling achieves considerably lower MSEs than additive modeling in  \cite{yang2018backfitting}.

\begin{table}[t]
  \centering
  \caption{Prediction mean squared errors for real data}\label{tab:real-mse}
  \begin{tabular}{>{}c*{3}{c}}\hline
   & \multirow{1}{*}{\bfseries Ozone Data} & \multirow{1}{*}{\bfseries Boston Data} & \multirow{1}{*}{\bfseries Tecator Data}\\
\hline
    dPAM1, linear & 17.65 (0.20) & 13.24 (0.25) & 4.52 (0.10)\\
  $\mathrm{ATV}$, $m=1$ & 17.10 (0.25) & 11.78 (0.39) & 10.29 (0.59)\\
  $\mathrm{FTV}$, $m=1$ & 17.16 (0.22) & 14.23 (0.30) & 10.46 (0.39)\\
  $\mathrm{ATV}$, $m=2$ & 16.27 (0.12) & 9.01 (0.30) & {\bf 0.83 (0.03)}\\
  $\mathrm{FTV}$, $m=2$ & {\bf 15.63 (0.14)} & {\bf 8.56 (0.15)} & 1.00 (0.07)\\
  COSSO & 16.04 (0.06) & 9.89 (0.08) & 0.92 (0.02)\\
  MARS, \texttt{degree} $=1$ & 18.50 (0.28) & 15.50 (0.19) & 7.13 (0.34)\\
  MARS, \texttt{degree} $=2$ & 17.95 (0.46) & 13.65 (0.53) & 4.84 (1.37)\\
  MARS, \texttt{degree} $=3$ & 18.11 (0.68) & 14.41 (0.41) & 3.46 (0.15)\\
  GBM, \texttt{depth} $=1$ & 18.06 (0.14) & 13.17 (0.22)& 9.52 (0.22)\\
  GBM, \texttt{depth} $=2$ & 17.51 (0.21) & 10.41 (0.21)& 8.10 (0.15)\\
  GBM, \texttt{depth} $=3$ & 16.99 (0.09) & 9.70 (0.13)& 7.73 (0.27)\\
% RF, \texttt{ntree} $=50$ & 16.58 (0.19) & 10.27 (0.29) & 20.55 (0.31)\\
  RF, \texttt{ntree} $=100$ & 16.36 (0.18) & 11.43 (0.43) & 18.78 (0.67)\\
  RF, \texttt{ntree} $=200$ & 16.33 (0.20) & 10.78 (0.50) & 18.57 (0.70)\\
  RF, \texttt{ntree} $=400$ & 16.40 (0.11) & 10.21 (0.34) & 18.38 (0.35)\\
% RF, \texttt{ntree} $=800$ & 16.40 (0.12) & 10.61 (0.19) & 18.07 (0.76)\\
\hline
  \end{tabular}
  \end{table}

\subsection{Logistic ANOVA modeling}

\subsubsection{Simulation study}

The data $Y_i$ and $X_i = (X_{i1}, \ldots, X_{i,10})$ are generated such that $Y_i \sim  \mathrm{Bernoulli}(\mathrm{expit}(f(X_i)))$,
where $X_{ij} \sim \mathrm{Uniform}\, [0,1]$ independently for $j=1,\ldots, 10$, and
$f(x)$ is defined as in Section~\ref{sec:linear-simulation}, depending only on $(x_1,\ldots,x_4)$, except that $g_1, \ldots, g_4$ are centered: $\int_0^1 g_j(z) \,\dif z=1$ for $j=1,\ldots,4$.
We conduct $100$ repeated simulations, each with training and validation sets.
The tuning parameters are selected to minimize the logistic loss on the validation set.
The logistic loss, accuracy, and AUC are then estimated by Monte Carlo integration using $10,000$ test data points.
We increase the sample size to $n=500, 1000, 2000$ for more meaningful comparison, since $n$ from 100 to 400 seems insufficient to achieve reasonable estimation for logistic modeling.
The results about test error rates are summarized in Table~\ref{tab:log-simulated-accuracy}, and those about the test logistic loss and AUC are shown in Table~\ref{tab:log-simulated-logloss} and
\ref{tab:log-simulated-auc} in the Supplement.

From Table~\ref{tab:log-simulated-accuracy}, the two versions of our method using piecewise cross-linear splines achieve the best two performances in classification accuracy
among all methods studied. The same conclusion can also be drawn in terms of the logistic loss and AUC.

\begin{table}[t]
  \centering
  \caption{Classification error rates minus oracle (\%) for logistic ANOVA simulation}\label{tab:log-simulated-accuracy}
  \begin{tabular}{>{}c*{3}{c}}\hline
   & \multirow{1}{*}{\bfseries $n=500$} & \multirow{1}{*}{\bfseries $n=1000$} & \multirow{1}{*}{\bfseries $n=2000$}\\
\hline
  $\mathrm{ATV}$, $m=1$ & 5.54 (0.14) & 3.56 (0.09) & 2.21 (0.04) \\
 $\mathrm{FTV}$, $m=1$ & 5.94 (0.12) & 4.09 (0.08) & 2.61 (0.05) \\
  $\mathrm{ATV}$, $m=2$ & {\bf 4.50 (0.13)} & {\bf 2.73 (0.07)} & {\bf 1.55 (0.04)} \\
  $\mathrm{FTV}$, $m=2$ & 4.70 (0.11) & 3.01 (0.07) & 1.76 (0.04) \\
  GBM, \texttt{depth} $=1$  & 5.66 (0.11) & 4.08 (0.07) & 3.10 (0.04) \\
  GBM, \texttt{depth} $=2$  & 5.24 (0.12) & 3.63 (0.07) & 2.52 (0.04) \\
  GBM, \texttt{depth} $=3$  & 5.08 (0.11) & 3.54 (0.07) & 2.44 (0.04) \\
  RF, \texttt{ntree} $=100$  & 6.24 (0.09) & 5.02 (0.08) & 4.01 (0.05) \\
  RF, \texttt{ntree} $=200$  & 5.99 (0.09) & 4.75 (0.07) & 3.68 (0.05) \\
  RF, \texttt{ntree} $=400$  & 5.90 (0.09) & 4.61 (0.07) & 3.53 (0.05) \\
% RF, \texttt{ntree} $=800$  & 5.84 (0.09) & 4.53 (0.07) & 3.44 (0.05) \\
\hline
  \end{tabular}\\[.1in]
  \parbox{1\textwidth}{\small Note: The oracle error rate is $35.35\%$, estimated from classification using the true function $f(x)$.}
\end{table}

\subsubsection{Empirical examples}

We conduct experiments on four real datasets to evaluate our method. These datasets are available from UC Irvine Machine Learning Repository \url{https://archive.ics.uci.edu/ml/datasets.html} \citep{Dua2019}. The datasets are the BUPA Liver Disorder data, the PIMA Indian Diabetes data, the Australian Credit Approval data and the Diabetic Retinopathy Debrecen data. We pick these four datasets mainly because they have a reasonable number of input variables, compared with sample sizes, for investigating two-way interactions.
After dropping categorical variables with less than 4 levels in the Australian and Diabetic data, the number of input variables $p$ and sample size $n$ are listed in Table~\ref{tab:real-accuracy}.

For each dataset, we randomly select $2/3$ of the data for training with 10-fold cross validation, and the remaining $1/3$ of the data for testing. We repeat this process for 10 times and report the average of test performance with the standard error for each method.
Similarly as before, the hyper-parameters $\lambda$ and $\rho$ are tuned for our method, \texttt{shrinkage} and \texttt{n.trees} for \texttt{gbm}, and \texttt{mtry} for \texttt{randomForest} to minimize the logistic loss by cross validation. The results about test accuracy (i.e., $1-$test error) are summarized in Table~\ref{tab:real-accuracy}, and those about the test logistic loss and AUC are shown in Table~\ref{tab:real-auc} and \ref{tab:real-logloss} in the Supplement.

\begin{table}[H]
  \centering
  \caption{Classification accuracy rates (\%) for real data}\label{tab:real-accuracy}
  \begin{tabular}{>{}c*{4}{c}}\hline
   & \multirow{1}{*}{\bfseries BUPA} & \multirow{1}{*}{\bfseries Pima Indian} & \multirow{1}{*}{\bfseries Australian} & \multirow{1}{*}{\bfseries Diabetic}\\
\hline
$n$ & 345 & 768 & 680 & 1151 \\
$p$ & 6 & 8 & 8 & 16 \\
\hline
    dPAM1, linear  & 72.43 (0.37) & 76.25 (0.41) & 80.83 (0.60) & {\bf 75.70 (0.54)} \\
  $\mathrm{ATV}$, $m=1$ & 69.39 (2.16) & 75.35 (0.73) & 79.52 (0.86) & 65.63 (0.53)\\
  $\mathrm{FTV}$, $m=1$ & 70.26 (1.54) & 75.20 (0.69) & 78.52 (0.78) & 66.15 (0.52) \\
  $\mathrm{ATV}$, $m=2$ & {\bf 73.65 (0.61)} & {\bf 76.37 (0.66)} & 80.00 (0.69)& 74.43 (0.61) \\
  $\mathrm{FTV}$, $m=2$ & 73.04 (0.61) & 75.90 (0.72) & 79.04 (0.60) & 74.87 (0.42) \\
  GBM, \texttt{depth} $=1$ & 70.70 (0.86) & 75.27 (0.59) & 80.91 (0.76) & 69.38 (0.93) \\
  GBM, \texttt{depth} $=2$ & 70.70 (1.00) & 75.16 (0.53) & 81.17 (0.34) & 69.04 (0.75) \\
  GBM, \texttt{depth} $=3$ & 71.48 (1.19) & 75.90 (0.59) & 80.17 (0.82) & 69.24 (0.92) \\
% RF, \texttt{ntree} $=50$ & 71.57 (1.15) & 73.87 (0.59) & 80.39 (0.68) & 67.34 (0.79) \\
  RF, \texttt{ntree} $=100$ & 72.78 (1.36) & 75.20 (0.51) & 79.22 (0.76) & 65.96 (0.50) \\
  RF, \texttt{ntree} $=200$ & 71.22 (0.99) & 76.17 (0.65) & 80.22 (0.62) & 67.37 (0.69) \\
  RF, \texttt{ntree} $=400$ & 71.91 (1.05) & 75.59 (0.51) & {\bf 82.17 (0.66)} & 66.69 (0.71) \\
\hline
  \end{tabular}
  \end{table}

From Table~\ref{tab:real-accuracy}, the two versions of our method using piecewise cross-linear splines consistently achieve performances among the the best three, compared
with all versions of GBM and random forest, on the BUPA, Pima Indian, and Diabetic data.
For these three datasets, doubly penalized additive modeling in \cite{yang2018backfitting} already performs better than or similarly as the best
from all versions of GBM and random forest as well as our ANOVA modeling.
Our method with $m=1$ or 2 achieves competitive performances on the Australia data.
The results of GMB and random forest on the BUPA and Pima Indian data are consistent with those in \cite{breiman2001random}. Previous experiments on Diabetic data can be found in \cite{somu2016hypergraph}. Since we removed categorical variables with less than four levels, our results for Australian and Diabetic datasets can be different from previously reported results.

\section{Conclusion}\label{sec:conclusion}

We formulate hierarchical total variations and derive suitable basis functions for multivariate splines, so as to extend the backfitting algorithm in \cite{yang2018backfitting}
for doubly penalized ANOVA modeling. This is the first time that ANOVA modeling with multivariate total-variation penalties is developed for nonparametric regression.
The results from our numerical experiments are very encouraging and demonstrate considerable gains from our method.
Nevertheless, various questions remain to be fully addressed. It is desirable to develop algorithms for handling higher-order interactions,
investigate choices of tuning parameters $(\rho_k,\lambda_k)$ depending on $k$, and study comparison and combination with greedy methods.

\section{Appendix}\label{sec:appendix}

\subsection{Proof of Proposition~\ref{prop:additive}}
%\begin{proof}
For notational simplicity, the dependency on $(\rho_1,\ldots,\rho_d)$ is  suppressed.
For $m=1$, we have by Definition \ref{def:generalized-total-variation} and (\ref{eqn:anova-fk}),
\begin{align*}
   \mathrm{HTV}_d^{1}\big(g\big) &= \sum_{k=1}^d\sum_{S_k:|S_k|=k}\rho_{k}\mathrm{TV}_k\Big(\prod_{j \not\in S_k}H_j g\Big) \\
   &= \sum_{k=1}^d\sum_{S_k:|S_k|=k}\rho_{k}\mathrm{TV}_k\Big(\prod_{j\in S_k}\big(1-H_j\big)\prod_{j \not\in S_k}H_j g\Big)
   = \sum_{k=1}^d\sum_{S_k:|S_k|=k}\rho_{k}\mathrm{TV}_k\Big(g_{S_k}\Big).
\end{align*}
The second equality holds because $\prod_{j\in S_k}(1-H_j) h$ for a $k$-variate function $h$ is a sum of $h$ and another function in less than $k$ variables, the latter of which
does not affect the raw TV. For $m \ge 2$, we have by Definition \ref{def:generalized-total-variation},
  \begin{eqnarray*}
  \mathrm{HTV}_d^{m}\big(g\big) &=& \sum_{k=1}^d\sum_{S_k:|S_k|=k}\mathrm{HTV}_k^{m-1}\Big(\prod_{j \not\in S_k}H_j \prod_{j \in S_k}D_j g\Big) \\
   &=& \sum_{k=1}^d\sum_{S_k:|S_k|=k}\mathrm{HTV}_k^{m-1}\Big(\prod_{j \in S_k}D_j g_{S_k}\Big).
\end{eqnarray*}
The second equality holds because
\[
\prod_{j \in S_k}D_j g_{S_k} = \prod_{j \in S_k}D_j\prod_{j \in S_k}(1-H_j)\prod_{l\not\in S_k}H_j g = \prod_{l\not\in S_k}H_j \prod_{j \in S_k}D_j g,
\]
by (\ref{eqn:anova-fk}) and the fact that $H_jh$ is constant in $z_j$ and hence $D_j H_j h=0$ for any function $h$.
%\end{proof}

\subsection{Proof of Proposition~\ref{prop:lasso}}  %\label{prf:lemma}

First, we show that
\begin{align}
\mathrm{TV}_k\big(\beta^\T_{S_k}\Phi_{S_k}^{(1)}\big)  = \|\beta_{S_k}\|_1. \label{eqn:d-variable-tv}
\end{align}
Without loss of generality, fix $S_k =\{1,\ldots,k\}$ and denote the elements in $\Phi_{S_k}^{(1)}$ as $\phi^{(1)}_{\nu_1,\ldots,\nu_k}$ and those in $\beta_{S_k}$ as $b_{\nu_1,\ldots,\nu_k}$
for $2 \le \nu_1 \le n_1, \ldots, 2 \le \nu_k \le n_k$.
By Definition~\ref{def:multivariate-tv}, we have
\begin{eqnarray*}
  \mathrm{TV}_k\big(\beta^\T_{S_k}\Phi_{S_k}\big) &=& \sum_{i_1=1}^{n_1-1}\cdots\sum_{i_k=1}^{n_k-1}
  \Big|\sum_{\nu_1=2}^{n_1}\cdots\sum_{\nu_k=2}^{n_k} b_{\nu_1, \ldots, \nu_k}\Big\{\phi^{(1)}_{\nu_1,\ldots,\nu_k} \big(z_{i_1+1,1}, \ldots, z_{i_k+1,k}\big) \\
  &+& \sum_{\nu=1}^k (-1)^\nu \sum_{1\le j_1< \ldots < j_\nu \le k}
  \phi^{(1)}_{\nu_1,\ldots,\nu_k} \big(z_{i_1+1,1}, \ldots, z_{i_{j_1},j_1}, \ldots,z_{i_{j_l},j_l} \ldots, z_{i_k+1,k}\big)\Big\}\Big|  \\
  &=& \sum_{i_1=1}^{n_1-1}\cdots\sum_{i_k=1}^{n_k-1} \Big|b_{i_1+1, \ldots, i_k+1}\Big|  = \|\beta_{S_k}\|_1.
\end{eqnarray*}
The second equality holds because the quantity in the curly bracket above
equals $1$ if $\nu_1 = i_1 + 1, \ldots, \nu_k = i_k + 1$, and equals $0$ otherwise.

Next, we show by induction that for $g(z_1, \ldots, z_d) = \beta^\T_0 + \sum_{k=1}^d \sum_{S_k: |S_k|=k} \beta^\T_{S_k}\Psi_{S_k}^{(m)}$,
Eqn (\ref{eqn:HTV-lasso}) holds, that is, $\mathrm{HTV}^{m}_d (g) = \sum_{k=1}^d \sum_{S_k: |S_k|=k} \| R_{S_k}^{(m)} \beta_{S_k} \|_1$.
For $m=1$, we have by Proposition~\ref{prop:additive} and the fact that $g_{S_k} = \beta^\T_{S_k}\Psi_{S_k}^{(1)}$,
\begin{eqnarray*}
  \mathrm{HTV}_d^1\big(g \big) &=& \sum_{k=1}^d\sum_{S_k: |S_k|=k}\rho_k\mathrm{TV}_k\Big(g_{S_k}\Big) =\sum_{k=1}^d\sum_{S_k: |S_k|=k}\rho_k\mathrm{TV}_k\Big(\beta^\T_{S_k}\Psi_{S_k}^{(1)} \Big) \\
   &=& \sum_{k=1}^d\sum_{S_k: |S_k|=k}\rho_k \|\beta_{S_k}\|_1 .
\end{eqnarray*}
The last equality holds because $\mathrm{TV}_k\big(\beta^\T_{S_k}\Psi_{S_k}^{(1)} \big)= \mathrm{TV}_k\big(\beta^\T_{S_k}\Phi_{S_k}^{(1)}\big)$ and Eqn (\ref{eqn:d-variable-tv}).
This yields (\ref{eqn:HTV-lasso}) for $m=1$, because the non-differentiation degree of each basis function in $\Psi^{(1)}_{S_k}$ is $k$ and hence each diagonal element of $R_{S_k}^{(1)}$ is $\rho_k$.

Suppose that (\ref{eqn:HTV-lasso}) holds for $m=M-1$ with $M \ge 2$. For $m=M$, application of Proposition \ref{prop:additive} and the fact that $g_{S_k} = \beta^\T_{S_k}\Psi_{S_k}^{(M)}$ yields
\begin{eqnarray}
  \mathrm{HTV}_d^{M}\big(g \big) &=& \sum_{k=1}^d\sum_{S_k: |S_k|=k}\mathrm{HTV}^{M-1}_k\Big(\prod_{j\in S_k}D_j g_{S_k} \Big) \nonumber \\
  &=& \sum_{k=1}^d\sum_{S_k: |S_k|=k}\mathrm{HTV}^{M-1}_k\Big(\beta^\T_{S_k}\prod_{j\in S_k}D_j \Psi_{S_k}^{(M)} \Big) \nonumber \\
  &=& \sum_{k=1}^d\sum_{S_k: |S_k|=k}\mathrm{HTV}^{M-1}_k\Big(\beta^\T_{S_k}\prod_{j\in S_k}D_j \tilde \Phi_{S_k}^{(M)} \Big)  . \label{eqn:HTV-expression}
\end{eqnarray}
The last equality holds because by (\ref{eqn:psi-def2}),
\[
\prod_{j\in S_k}D_j\Psi_{S_k}^{(M)} =  \prod_{j\in S_k}D_j\prod_{j\in S_k}(1-H_j) \tilde \Phi_{S_k}^{(M)} = \prod_{j\in S_k}D_j \tilde \Phi_{S_k}^{(M)} .
\]

For further simplification, notice that by the recursive property (\ref{eqn:recursive}), for any $S_k=\{j_1,\ldots,j_k\}$ and $2 \le \nu_{j_1} \le n_{j_1}, ...,
2 \le \nu_{j_k} \le n_{j_k}$,
\begin{align}
\prod_{l=1}^k D_{j_l} \tilde \phi_{\nu_{j_l},j_l}^{(M)} = \prod_{1 \le l \le k: \, \nu_{j_l} \ge 3} (1- H_{j_l}) T^{(M-1)}_{j_l} D_{j_l} \phi^{(M)}_{\nu_{j_l},j_l}, \label{eqn:basis-mapping}
\end{align}
where $\tilde \phi_{\nu_{j_l},j_l}^{(M)} = T_{j_l}^{(M)} \phi_{\nu_{j_l},j_l}^{(M)} $.
Because $ \prod_{1 \le l \le k: \, \nu_{j_l} \ge 3} D_{j_l} \phi^{(M)}_{\nu_{j_l},j_l}$ is an element in $\Phi^{(M-1)}_{S_{k^\prime}}$
for $S_{k^\prime} = \{j_l: \nu_{j_l} \ge 3, 1\le l \le k\}$, Eqn (\ref{eqn:basis-mapping}) indicates that
each element in $\prod_{j\in S_k}D_j \tilde \Phi_{S_k}^{(M)}$ coincides with
exactly one element in $\prod_{j \in S_{k^\prime}} (1-H_j) T_j^{(M-1)} \Phi^{(M-1)}_{S_{k^\prime}}$ which is $\Psi_{S_{k^\prime}}^{(M-1)} $ by (\ref{eqn:psi-def2}),
except that the basis $\prod_{j\in S_k} z_j$ in $\tilde \Phi_{S_k}^{(M)}$ yields the constant 1 after the cross-differentiation $\prod_{j\in S_k}D_j$.
Hence $\beta^\T_{S_k} \prod_{j\in S_k}D_j \tilde \Phi_{S_k}^{(M)}$ can be written as
\begin{align}
\beta^\T_{S_k}\prod_{j\in S_k}D_j \tilde \Phi_{S_k}^{(M)} = \beta_0^\prime + \sum_{S_{k^\prime} \subset S_k }
\beta^{\prime^\T}_{S_{k^\prime}} \Psi_{S_{k^\prime}}^{(M-1)} , \label{eqn:basis-link}
\end{align}
where each nonzero element in $\beta_{S_k}$ corresponds to exactly one nonzero element in $\beta^{\prime}_0$ and $\{ \beta^{\prime}_{S_{k^\prime}}: S_{k^\prime} \subset S_k\}$ and vice versa.
Moreover, the non-differentiation degree of a basis function in $\Psi^{(M)}_{S_k}$ is not affected by the cross-differentiation $\prod_{j\in S_k}D_j$ and hence
is the same as that of the corresponding basis function in $\{\Psi_{S_{k^\prime}}^{(M-1)} : S_{k^\prime} \subset S_k\}$ by (\ref{eqn:basis-mapping}). Hence
\begin{align}
\| R_{S_k}^{(M)} \beta^\T_{S_k} \|_1 = \sum_{S_{k^\prime} \subset S_k } \| R_{S_{k^\prime}}^{(M-1)} \beta^{\prime^\T}_{S_{k^\prime}} \|_1 . \label{eqn:norm-mapping}
\end{align}
Then (\ref{eqn:HTV-lasso}) holds for $m=M$ by (\ref{eqn:HTV-expression}), (\ref{eqn:norm-mapping}), and the fact that
$\mathrm{HTV}^{M-1}_k(\beta^\T_{S_k}\prod_{j\in S_k}D_j \tilde\Phi_{S_k}^{(M)})$ is $\mathrm{HTV}^{M-1}_k$ of the right-hand side of (\ref{eqn:basis-link}), which by induction is
$\sum_{S_{k^\prime} \subset S_k } \| R_{S_{k^\prime}}^{(M-1)} \beta^{\prime}_{S_{k^\prime}} \|_1 $.

\bibliography{reference_new-05042019}
\bibliographystyle{myapa}   %Heng

\clearpage

\setcounter{page}{1}

\setcounter{section}{0}
\setcounter{equation}{0}

\renewcommand{\theequation}{S\arabic{equation}}
\renewcommand{\thesection}{S\arabic{section}}

\setcounter{table}{0}
\setcounter{figure}{0}

\renewcommand\thefigure{S\arabic{figure}}
\renewcommand\thetable{S\arabic{table}}

\begin{center}
{\Large Supplementary Material for \\``Hierarchical Total Variations and Doubly Penalized ANOVA Modeling for Multivariate Nonparametric Regression"}

\vspace{.1in} Ting Yang and  Zhiqiang Tan
\end{center}

\section{2-dimensional smoothing} \label{sec:2d-smooth}

\begin{figure}[H]
  \centering
  \includegraphics[width=2in, height=2in]{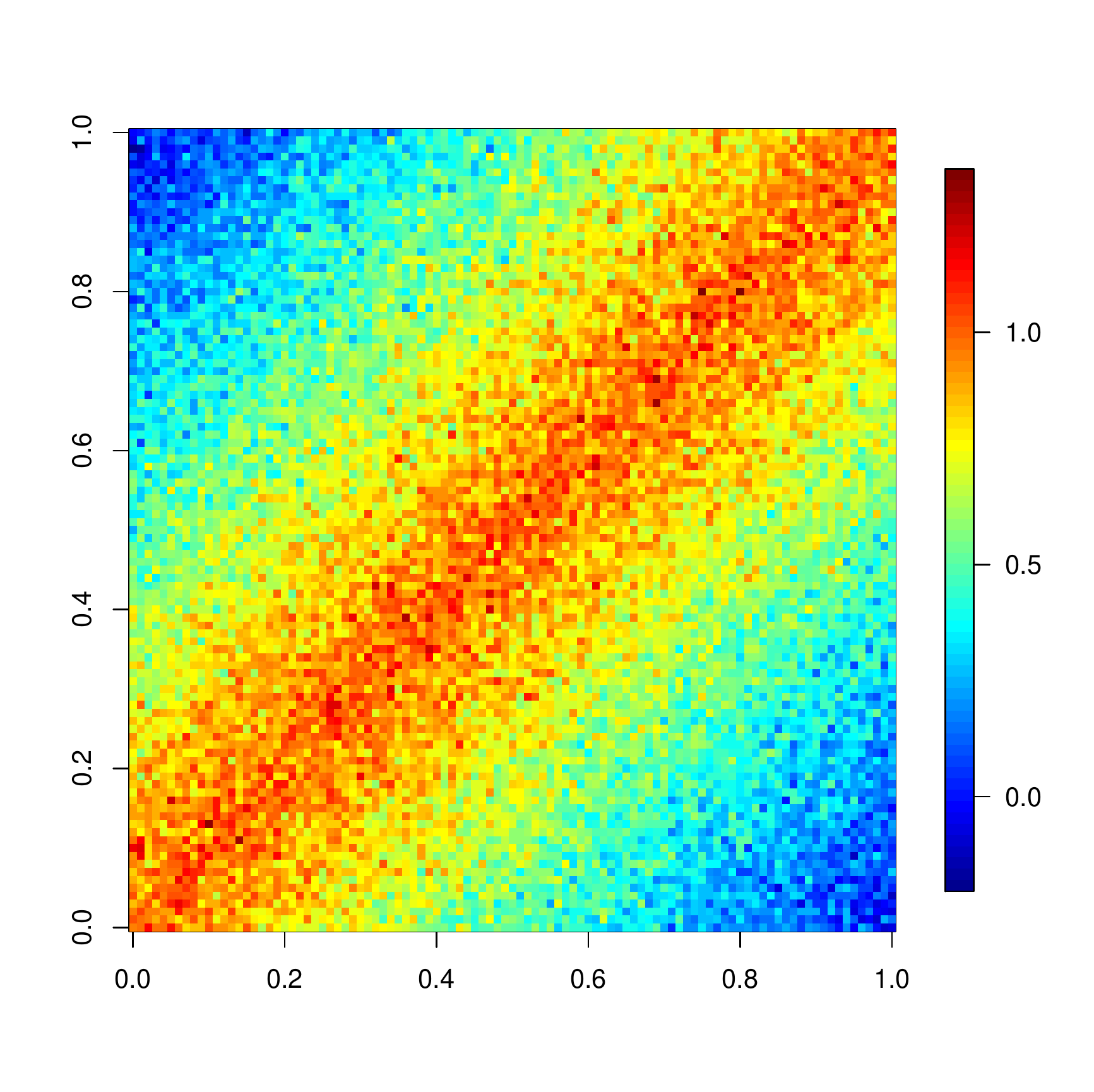}\\
  \caption{True function in 2-dimensional smoothing.}\label{fig:fixed-point}
\end{figure}

To compare the use of fixed-point and averaging TV as two versions of hierarchical TV, we conduct a simple experiment on 2-dimensional smoothing.
We randomly sample pairs $\{(X_{1i}, X_{2i}): i = 1, 2, \ldots, 100\}$ from the uniform distribution on $[0, 1] \times [0,1]$ and then generate
$Y_i = f(X_{1i}, X_{2i}) + \epsilon_i$, where $\epsilon_i \sim \mathrm{N}(0, 0.1^2)$ and
\[
f(x_1, x_2) = 1 - |x_1 - x_2| .
\]
The true function $f(x)$ is plotted in Figure~\ref{fig:fixed-point} on a $101\times 101$ equally spaced grid.

We implement doubly penalized least-squares estimation with $p=K=2$ using either the fixed-point TV or averaging TV penalty.
For fixed-point TV, the fixed points used are $(x_{\min}, y_{\min})$, $(x_{\max}, y_{\max})$, $(x_{\min}, y_{\max})$, $(x_{\max}, y_{\min})$, $(x_{\mathrm{median}}, y_{\mathrm{median}})$.
The marginal knots are 11 quantiles by 10\%. Both TV and empirical-norm penalties are tuned on a simulated validation dataset, similarly as in Section~\ref{sec:linear-simulation}.
To evaluate the performance, the MISE is calculated by averaging the squared errors at the $101^2$ grid points.
The simulation is repeated for 100 times. The means and standard errors of MISEs are reported in Table~\ref{tab:toy-mise}.

\begin{table}[H]
  \centering
  \caption{MISEs in 2-dimensional smoothing}\label{tab:toy-mise}
  \begin{tabular}{>{}c*{2}{c}}\hline
 \multirow{1}{*}{\bfseries $(x_{\min}, y_{\min})$}  & \multirow{1}{*}{\bfseries $(x_{\max}, y_{\max})$} & \multirow{1}{*}{\bfseries $(x_{\min}, y_{\max})$} \\
  $0.0050 \,(1.1\times 10^{-4})$ & $0.0040 \,(8\times 10^{-5})$ & $0.0046 \,(1.0 \times 10^{-4})$  \\
  \hline
  \multirow{1}{*}{\bfseries $(x_{\max}, y_{\min})$} & \multirow{1}{*}{\bfseries $(x_{\mathrm{median}}, y_{\mathrm{median}})$} & \multirow{1}{*}{ Averaged }\\
 $0.0046 \,(1.3\times 10^{-4})$ & $0.0048 \,(1.1 \times 10^{-4})$ & $0.0040 \,(9.9\times10^{-5})$   \\
\hline
  \end{tabular}
  \end{table}

\begin{figure}[H]
  \centering
  \includegraphics[width=6in, height=4in]{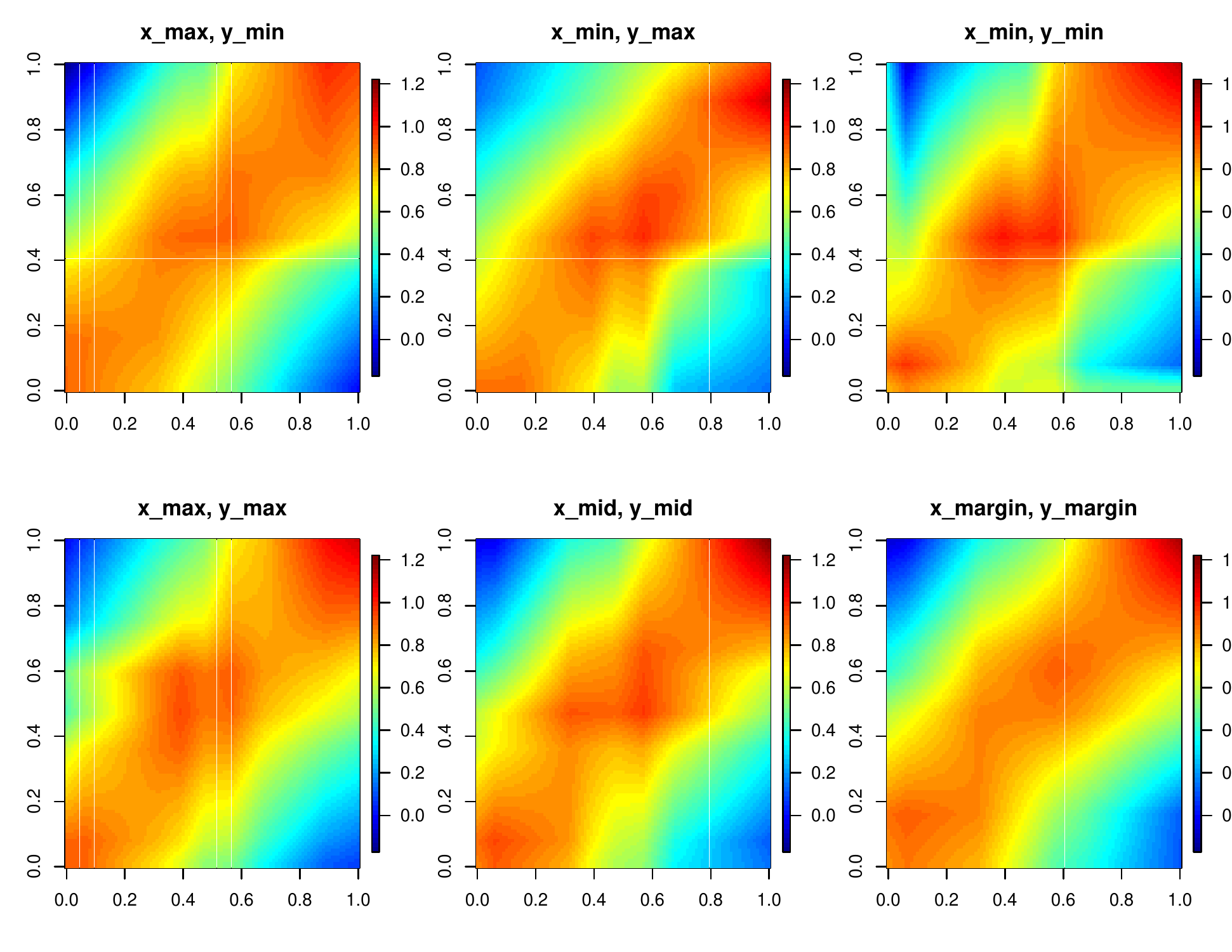}\\
  \caption{Estimated functions in 2-dimensional smoothing.}\label{fig:2d-toy-visual}
\end{figure}

From Table~\ref{tab:toy-mise}, the performance from using fixed-point TVs depends on the choice of the fixed point.
The estimated functions are plotted on the $101\times 101$ grid in Figure~\ref{fig:2d-toy-visual}.
We see that the functions estimated by using fixed-point TVs have more artificial stripes and corners than that estimated using the averaging TV,
which seems much smoother.


\section{Additional numerical results}

\subsection{Simulated study on linear ANOVA modeling}

\begin{figure}[H]
  \centering
  \includegraphics[width=5in, height=3.5in]{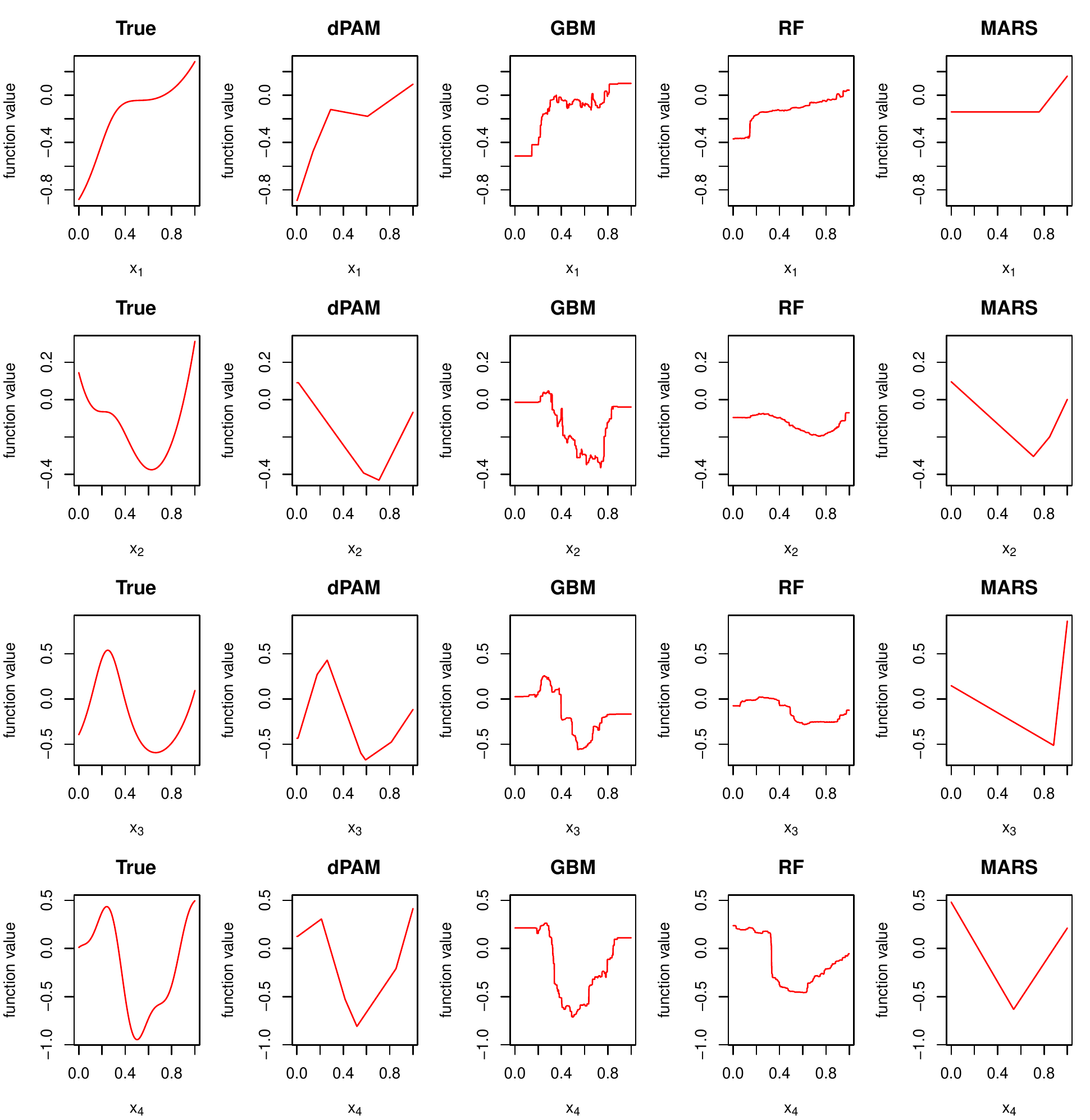}\\
  \caption{$1$-dimensional partial dependence plots for $p=10$ and $n=100$.}\label{fig:1d-partial-dependence-plot-n-100}
\end{figure}

\begin{figure}[H]
  \centering
  \includegraphics[width=5in, height=3in]{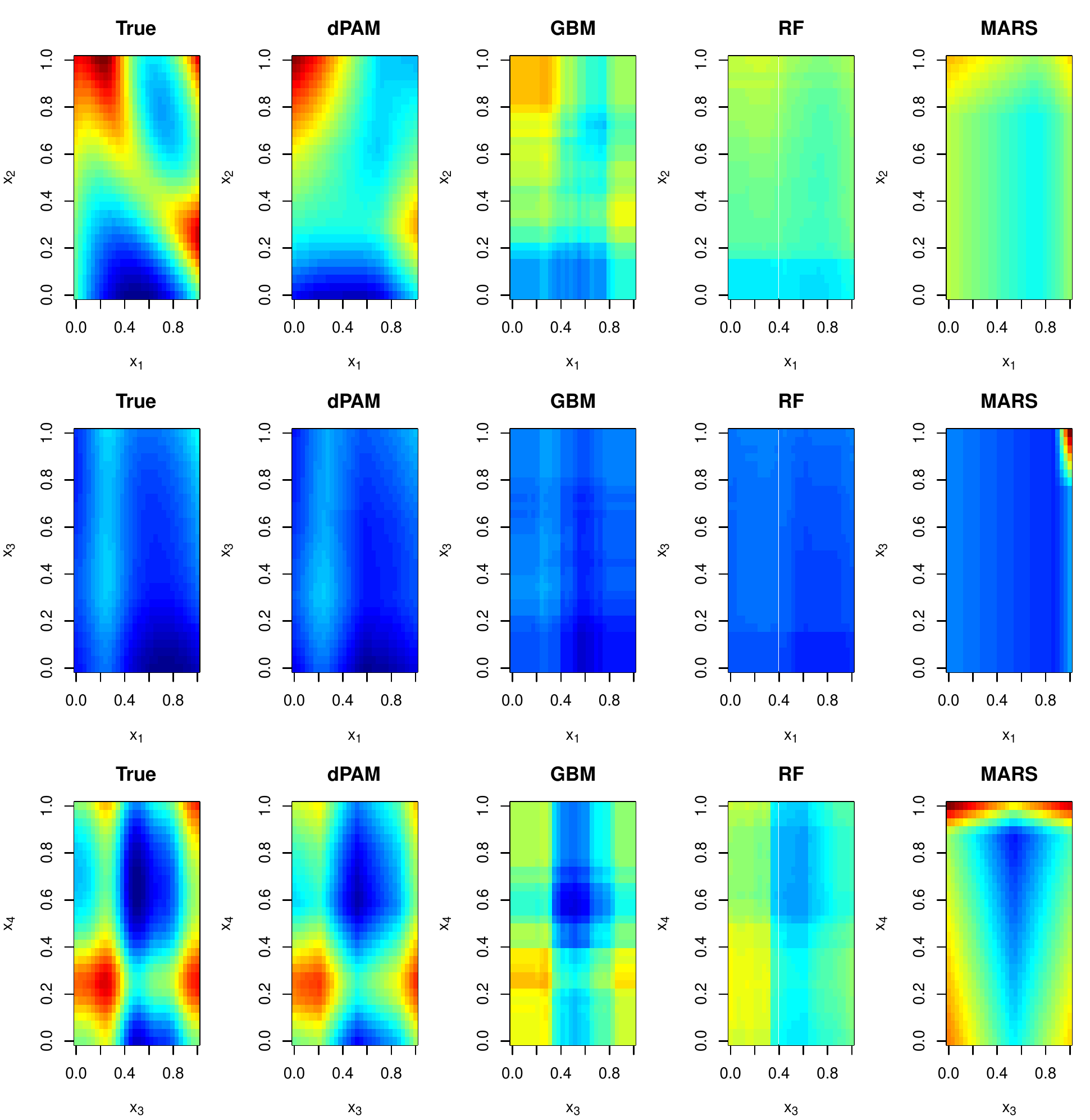}\\
  \caption{$2$-dimensional partial dependence plots for $p=10$ and $n=100$.}\label{fig:2d-partial-dependence-plot-n-100}
\end{figure}

\begin{figure}[H]
  \centering
  \includegraphics[width=5in, height=3.5in]{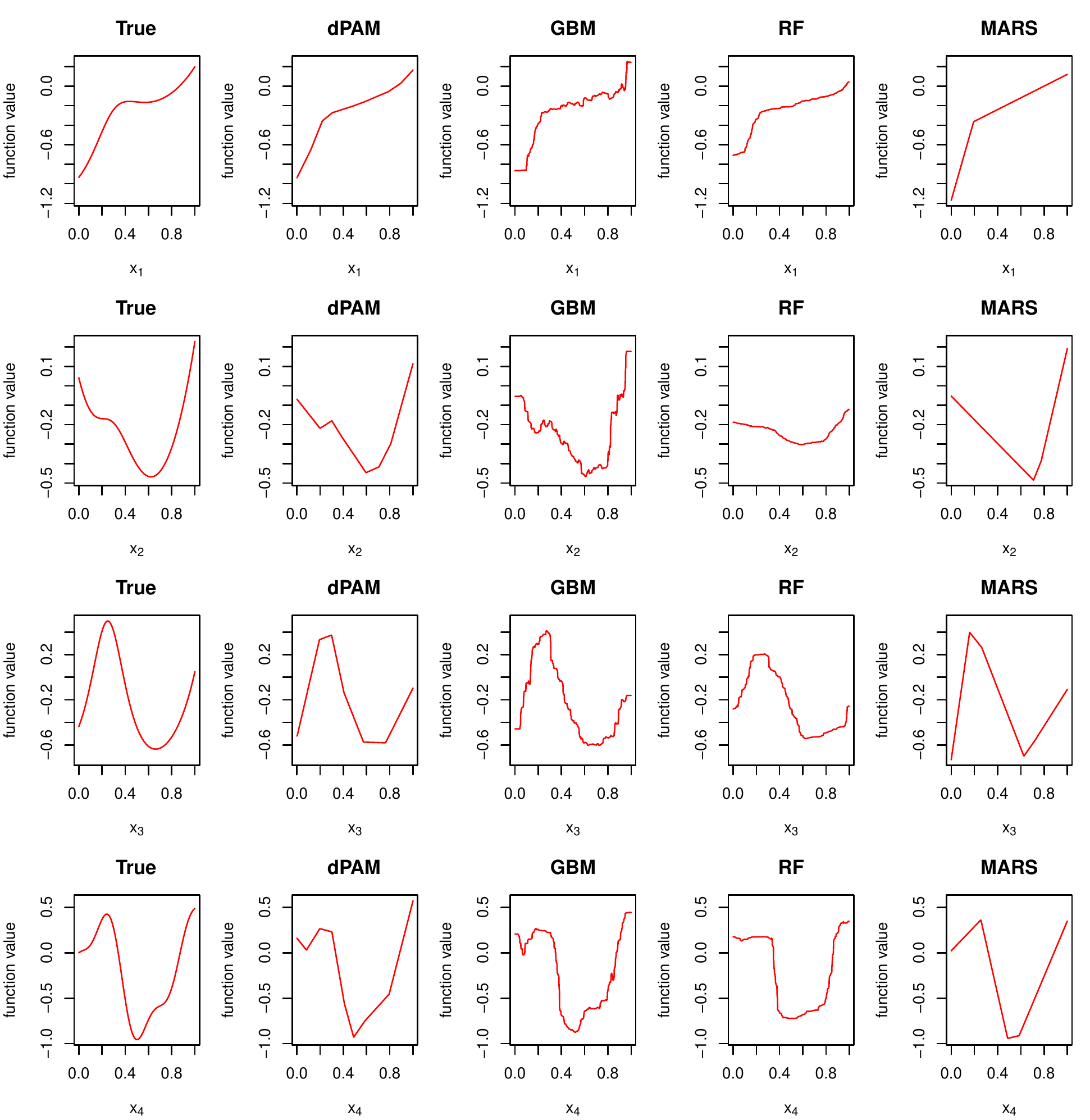}\\
  \caption{$1$-dimensional partial dependence plots for $p=10$ and $n=400$.}\label{fig:1d-partial-dependence-plot-n-400}
\end{figure}

\begin{figure}[H]
  \centering
  \includegraphics[width=5in, height=3in]{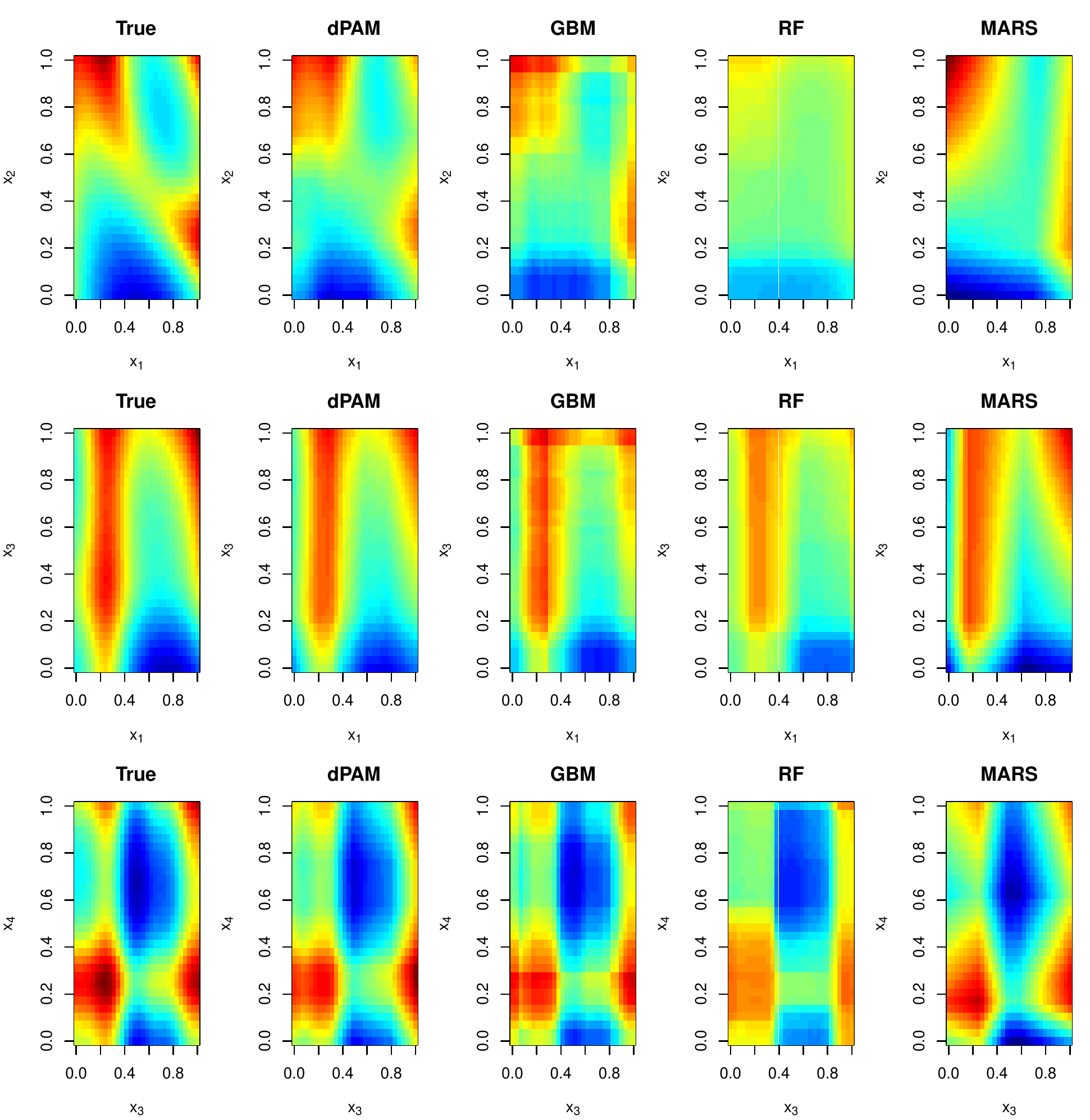}\\
  \caption{$2$-dimensional partial dependence plots for $p=10$ and $n=400$.}\label{fig:2d-partial-dependence-plot-n-400}
\end{figure}

\subsection{Simulated study on logistic ANOVA modeling}

\begin{table}[H]
  \centering
  \caption{Logistic losses minus oracle for logistic simulation}\label{tab:log-simulated-logloss}
  \begin{tabular}{>{}c*{3}{c}}\hline
   & \multirow{1}{*}{\bfseries $n=500$} & \multirow{1}{*}{\bfseries $n=1000$} & \multirow{1}{*}{\bfseries $n=2000$}\\
\hline
  $\mathrm{ATV}$, $m=1$ & 0.0412 (0.0007) & 0.0292 (0.0005) & 0.0199 (0.0003) \\
  $\mathrm{FTV}$, $m=1$& 0.0436 (0.0006) & 0.0328 (0.0005) & 0.0227 (0.0003) \\
  $\mathrm{ATV}$, $m=2$ & {\bf 0.0352 (0.0008)} & {\bf 0.0235 (0.0004)} & {\bf 0.0151 (0.0003)} \\
  $\mathrm{FTV}$, $m=1$ & 0.0373 (0.0006) & 0.0252 (0.0004) & 0.0160 (0.0003) \\
  GBM, \texttt{depth} $=1$ & 0.0425 (0.0006) & 0.0329 (0.0004) & 0.0263 (0.0002) \\
  GBM, \texttt{depth} $=2$ & 0.0401 (0.0006) & 0.0304 (0.0004) & 0.0221 (0.0003) \\
  GBM, \texttt{depth} $=3$ & 0.0392 (0.0005) & 0.0295 (0.0005) & 0.0214 (0.0003) \\
  RF, \texttt{ntree} $=100$ & 0.0492 (0.0005) & 0.0408 (0.0004) & 0.0332 (0.0003) \\
  RF, \texttt{ntree} $=200$ & 0.0465 (0.0006) & 0.0377 (0.0004) & 0.0308 (0.0003) \\
  RF, \texttt{ntree} $=400$ & 0.0452 (0.0005) & 0.0363 (0.0004) & 0.0292 (0.0003) \\
% RF, \texttt{ntree} $=800$ & 0.0441 (0.0005) & 0.0356 (0.0004) & 0.0286 (0.0002) \\
\hline
  \end{tabular}\\[.1in]
  \parbox{1\textwidth}{\small Note: The oracle logistic loss is $0.6276$, estimated using the true function $f(x)$.}
  \end{table}

\begin{table}[H]
  \centering
  \caption{Oracle minus achieved AUC (\%) for logistic simulation}\label{tab:log-simulated-auc}
  \begin{tabular}{>{}c*{3}{c}}\hline
   & \multirow{1}{*}{\bfseries $n=500$} & \multirow{1}{*}{\bfseries $n=1000$} & \multirow{1}{*}{\bfseries $n=2000$}\\
\hline
  $\mathrm{ATV}$, $m=1$ & 7.49 (0.18) & 4.75 (0.10) & 2.96 (0.05) \\
  $\mathrm{FTV}$, $m=1$ & 8.06 (0.15) & 5.64 (0.10) & 3.68 (0.06) \\
  $\mathrm{ATV}$, $m=2$ & {\bf 6.02 (0.17)} & {\bf 3.65 (0.07)} & {\bf 2.19 (0.05)} \\
  $\mathrm{FTV}$, $m=2$ & 6.42 (0.13) & 4.13 (0.08) & 2.45 (0.05) \\
  GBM, \texttt{depth} $=1$  & 7.58 (0.12) & 5.65 (0.09) & 4.34 (0.05) \\
  GBM, \texttt{depth} $=2$  & 6.97 (0.12) & 5.10 (0.09) & 3.51 (0.05) \\
  GBM, \texttt{depth} $=3$  & 6.77 (0.12) & 4.90 (0.09) & 3.37 (0.05) \\
  RF, \texttt{ntree} $=100$  & 8.77 (0.12) & 7.11 (0.09) & 5.74 (0.06) \\
  RF, \texttt{ntree} $=200$  & 8.39 (0.12) & 6.70 (0.09) & 5.31 (0.06) \\
  RF, \texttt{ntree} $=400$  & 8.23 (0.11) & 6.48 (0.09) & 5.06 (0.05) \\
% RF, \texttt{ntree} $=800$  & 8.12 (0.11) & 6.37 (0.09) & 4.94 (0.05) \\
\hline
  \end{tabular}\\[.1in]
  \parbox{1\textwidth}{\small Note: The oracle AUC is $69.79\%$, estimated using the true function $f(x)$.}
  \end{table}

\subsection{Empirical examples on logistic ANOVA modeling}

  \begin{table}[H]
  \centering
  \caption{Logistic losses for real data}\label{tab:real-logloss}
  \begin{tabular}{>{}c*{4}{c}}\hline
   & \multirow{1}{*}{\bfseries BUPA} & \multirow{1}{*}{\bfseries Pima Indian} & \multirow{1}{*}{\bfseries Australian} & \multirow{1}{*}{\bfseries Diabetic} \\
\hline
$n$ & 345 & 768 & 680 & 1151 \\
$p$ & 6 & 8 & 8 & 16 \\
\hline
    dPAM1, linear  & {\bf 0.553 (0.010)} & 0.482 (0.008) & 0.434 (0.011) & {\bf 0.488 (0.007)} \\
 $\mathrm{ATV}$, $m=1$ & 0.585 (0.016) & 0.493 (0.009) & 0.458 (0.011) & 0.610 (0.003) \\
 $\mathrm{FTV}$, $m=1$ & 0.594 (0.017) & 0.499 (0.010) & 0.460 (0.012) & 0.597 (0.006) \\
  $\mathrm{ATV}$, $m=2$ & 0.558 (0.010) & 0.492 (0.010) & 0.456 (0.012) & 0.500 (0.008) \\
  $\mathrm{FTV}$, $m=2$ & 0.556 (0.009) & 0.489 (0.009) & 0.456 (0.008) & 0.508 (0.014) \\
  GBM, \texttt{depth} $=1$ & 0.583 (0.006) & {\bf 0.479 (0.008)} & 0.423 (0.013) & 0.567 (0.007) \\
  GBM, \texttt{depth} $=2$ & 0.572 (0.010) & 0.496 (0.007) & {\bf 0.410 (0.006)} & 0.581 (0.009) \\
  GBM, \texttt{depth} $=3$ & 0.568 (0.009) & 0.482 (0.008) & 0.441 (0.012) & 0.582 (0.008) \\
% RF, \texttt{ntree} $=50$ & 0.580 (0.008) & 0.508 (0.009) & 0.500 (0.022) & 0.607 (0.011) \\
  RF, \texttt{ntree} $=100$ & 0.582 (0.008) & 0.497 (0.010) & 0.492 (0.021) & 0.597 (0.005) \\
  RF, \texttt{ntree} $=200$ & 0.596 (0.013) & 0.492 (0.007) & 0.464 (0.015) & 0.591 (0.005) \\
  RF, \texttt{ntree} $=400$ & 0.584 (0.008) & 0.493 (0.005) & 0.437 (0.006) & 0.594 (0.005) \\
\hline
  \end{tabular}
  \end{table}

 \begin{table}[H]
  \centering
  \caption{Achieved AUC (\%) for real data}\label{tab:real-auc}
  \begin{tabular}{>{}c*{4}{c}}\hline
   & \multirow{1}{*}{\bfseries BUPA} & \multirow{1}{*}{\bfseries Pima Indian} & \multirow{1}{*}{\bfseries Australian} & \multirow{1}{*}{\bfseries Diabetic} \\
\hline
$n$ & 345 & 768 & 680 & 1151 \\
$p$ & 6 & 8 & 8 & 16 \\
\hline
   dPAM1, linear  & {\bf 78.60 (0.88)} & {\bf 83.74 (0.57)} & 87.50 (0.78) & {\bf 83.40 (0.52)} \\
  $\mathrm{ATV}$, $m=1$ & 74.97 (1.65) & 82.67 (0.85) & 85.76 (0.79) & 72.21 (0.58)\\
  $\mathrm{FTV}$, $m=1$ & 75.23 (1.64) & 82.72 (0.80) & 85.75 (0.83) & 73.21 (0.74) \\
  $\mathrm{ATV}$, $m=2$ & 78.23 (0.75) & 83.48 (0.59) & 86.59 (0.63) & 82.55 (0.57)\\
  $\mathrm{FTV}$, $m=2$ & 78.37 (0.79) & 83.65 (0.55) & 86.50 (0.59) & 82.73 (0.54) \\
  GBM, \texttt{depth} $=1$ & 75.80 (0.74) & 83.25 (0.73) & 88.58 (0.69) & 76.87 (0.67) \\
  GBM, \texttt{depth} $=2$ & 76.02 (1.06) & 81.88 (0.64) & {\bf 88.93 (0.38)} & 76.00 (0.78)\\
  GBM, \texttt{depth} $=3$ & 76.23 (1.27) & 82.84 (0.73) & 87.33 (0.65) & 75.77 (0.83)\\
% RF, \texttt{ntree} $=50$ & 75.71 (0.98) & 81.14 (0.44) & 86.56 (0.41) & 73.98 (0.55)\\
  RF, \texttt{ntree} $=100$ & 77.29 (1.21) & 81.42 (0.39) & 86.20 (0.74) & 73.41 (0.68)\\
  RF, \texttt{ntree} $=200$ & 74.47 (1.58) & 82.36 (0.33) & 87.16 (0.57) & 74.49 (0.69)\\
  RF, \texttt{ntree} $=400$ & 75.74 (1.19) & 81.73 (0.45) & 88.02 (0.56) & 74.13 (0.59)\\
\hline
  \end{tabular}
  \end{table}

\end{document}